\documentclass{aa}
\usepackage{natbib}
\bibpunct{(}{)}{;}{a}{}{,} 

\usepackage[colorlinks=true, citecolor=blue, linkcolor=blue]{hyperref}
\usepackage{graphicx}
\usepackage{booktabs}
\usepackage{multirow}
\usepackage{tikz}
\usepackage{txfonts}
\usepackage{acronym}

\acrodef{PSF}{point spread function}
\acrodef{IFT}{Information Field Theory}
\acrodef{NIFTy}{Numerical Information Field Theory}
\acrodef{RL}{Richardson-Lucy}
\acrodef{S/N}{signal-to-noise ratio}
\acrodef{SSIM}{structural similarity index}
\acrodef{RMSE}{root mean squared error}
\acrodef{NLL}{negative log-likelihood}
\acrodef{MGVI}{Metric Gaussian Variational Inference}
\acrodef{MAP}{maximum-a-posteriori}
\acrodef{CALDB}{calibration database}
\acrodef{TV}{total variation}
\acrodef{FOV}{field of view}

\begin{document}
   \title{Bayesian Image Reconstruction with Spatially Variant PSFs \\ in X-ray Astronomy}
   \author{V. Eberle \inst{1, 2}
          \and
          M. Guardiani \inst{1, 2}
          \and
          M. Westerkamp \inst{1, 2}
          \and
          P. Arras \inst{1}
          \and
          P. Frank \inst{1, 5}
          \and
          J. Stadler \inst{1, 6}
          \and
          T. En\ss lin \inst{1, 2, 3, 4}
          }

   \institute{Max Planck Institute for Astrophysics, Karl-Schwarzschild-Straße 1, 85748 Garching, Germany,
        \and
            Faculty of Physics, Ludwig-Maximilians-Universität München, Geschwister-Scholl-Platz 1, 80539 München, Germany
        \and
            Deutsches Zentrum für Astrophysik, Postplatz 1, 02826 Görlitz, Germany
        \and 
            Excellence Cluster ORIGINS, Boltzmannstr. 2, 85748 Garching, Germany
        \and 
            Kavli Institute for Particle Astrophysics \& Cosmology, P.O. Box 2450, Stanford University, Stanford, CA 94395, USA 
        \and 
            University Observatory, Faculty of Physics, Ludwig-Maximilians-Universität, Scheinerstr. 1, 81679 Munich, Germany}

   \date{Received XXX, 2026}

\abstract{X-ray observatories introduce unwanted instrumental effects into photon count data that must be accurately accounted for. In particular, the spatially variant \ac{PSF} and shot noise pose a non-trivial inverse problem. We apply Bayesian inference grounded in \ac{IFT} combined with a fast spatially variant \ac{PSF} representation to infer the posterior mean and uncertainty of the X-ray flux under a minimal set of physically motivated prior assumptions, thereby removing or reducing these instrumental effects. 
First, ignoring spatial \ac{PSF} variability, the IFT-based deconvolution is benchmarked on a synthetic example against three variants of the \ac{RL} algorithm using SSIM, RMSE, and NLL as evaluation metrics.
Second, the ability of an interpolated patch-based convolution scheme to accurately represent the complex, spatially variant Chandra \ac{PSF} is assessed on simulated \ac{PSF} data. Finally, deblurring and spatially variant PSF representation are combined and applied to real Chandra observations of the supernova remnant Cassiopeia A. The IFT-based deconvolution outperforms all tested \ac{RL} variants on the synthetic benchmark. The patch-based convolution scheme accurately recovers the complex structure of the Chandra \ac{PSF}. Applied to Cassiopeia A, the reconstructed image appears visually sharper than the exposure-corrected data, and the data residuals are consistent with pure noise, indicating an accurate model of both the instrument and the sky. A comparison with a highly resolved reference dataset confirms this impression. The presented framework enables the removal of spatially variant \acp{PSF} from noisy data and opens avenues for future work on cross-calibration between multiple observations or different X-ray instruments, building on the principled uncertainty quantification inherent to the Bayesian framework.
}
\maketitle
\nolinenumbers 

\section{Introduction}
\acresetall
Optical imaging systems such as telescopes, microscopes, and mirrors are affected by aberrations and other image-degrading effects. 
These distortions significantly reduce spatial resolution and compromise the scientific value of the acquired data. 
While improving instrument design to minimize such effects is crucial, it is equally important to develop robust computational methods for correcting these degradations after data acquisition.
In the regime of high-\ac{S/N}, image deconvolution is well-established and often achievable with high fidelity.
However, as \ac{S/N} decreases, the inverse problem becomes increasingly ill-posed. 
Moreover, real-world instruments often exhibit spatially variant \acp{PSF} caused by optical aberrations such as spherical aberration or coma. 
Representing these \acp{PSF} requires high-dimensional models that are computationally expensive and memory-intensive, posing a major challenge for Bayesian imaging frameworks where efficiency and scalability are critical.
A further complication arises when the \ac{PSF} is not known precisely.
While information from the \acp{CALDB} or simulations can provide approximate \ac{PSF} models, they often contain uncertainties that must be accounted for in the reconstruction process.
Two core challenges therefore emerge: first, the robust deconvolution of noisy, low-signal data while preserving physical consistency and enabling uncertainty quantification; and second, the efficient, accurate, and scalable representation of spatially variant \acp{PSF}.

One of the most widely used deconvolution methods is \ac{RL} deconvolution \citep{richardson1972bayesian, lucy1974iterative}, which provides a maximum-likelihood solution under Poisson noise. 
However, for noisy data this results in noisy reconstructions due to overfitting, and the algorithm must be stopped at the right iteration, which in practice is difficult without knowledge of the ground truth.
Several regularized variants have been proposed to improve stability by adding regularization terms to the objective \citep{dey_richardsonlucy_2006}, but these require careful tuning of the regularization parameter, which is typically problem-specific.
If statistical knowledge about both the noise and the image is available, methods such as Wiener deconvolution can be applied.
In astrophysical X-ray imaging, where photon counts follow Poisson statistics and noise is both significant and well-understood, classical approaches, like the \ac{RL} algorithm, risk introducing artifacts. 

Turning to the representation of spatially variant \acp{PSF}, 
a compact review is given in \citet{thiebaut2016spatially}, whereas a comprehensive review can be found in \citet{denis2015fast}. 
Both review multiple approaches of representing spatially variant \acp{PSF} using interpolation and decomposition into basis functions. 
Furthermore, \citet{lauer2002deconvolution} shows that Karhunen-Loève transformation can be used to decompose the \ac{PSF} into orthogonal basis functions.
However, the methods discussed therein are mostly coupled with standard deconvolution approaches that focus on point-estimate solutions and therefore do not provide uncertainty quantification. 
In contrast, \citet{sakai2023richardson} incorporates uncertainty quantification by error propagation, while combining \ac{RL} deconvolution with a spatially variant \ac{PSF} assuming a monoenergetic PSF at $2.3\,\mathrm{keV}$.
A very different approach for spatially variant \ac{PSF} representation was developed in \citet{ramos2024solar}. Here, a U-Net is trained on simulated \acp{PSF} and used as an emulator for \ac{PSF} representation, which is particularly relevant in adaptive optics where the \ac{PSF} varies due to atmospheric turbulence. 
However, this approach is only validated in higher-\ac{S/N} regimes and does not provide a natural formulation for uncertainty quantification.
Alternative representations based on butterfly matrices, which exploit the structure of the fast Fourier transform, have also been proposed and showed promising first results \citep{e25040652, psf2022005033}.

To address the two core problems, the deconvolution and the spatially variant \ac{PSF} representation, we adopt a Bayesian framework grounded in \ac{IFT}~\citep{ensslin2009} and combine it with a spatially variant \ac{PSF} representation developed by \citet{nagy1997fast}. 
\ac{IFT} is implemented via the Python package \ac{NIFTy}~\citep{selig2013nifty, arras2019nifty5,niftyre} and enables the construction of generative models that naturally incorporate prior knowledge, handle uncertainties, and support variational inference, making it particularly suitable for inverse problems with complex noise and \ac{PSF} structures. 
\citet{selig2015denoising} developed an algorithm based on \ac{IFT} to denoise, deconvolve, and decompose photon observations and applied it to the Fermi $\gamma$-ray sky~\citep{selig2015denoised}. The algorithm was extended by \citet{pumpe2018denoising} to include information about spectral prior knowledge, and applied to more recent data of Fermi by \citet{scheel2023multicomponent}.
Beyond this, \ac{IFT}-based algorithms have found applications in X-ray astronomy, including the analysis of SN1006 \citep{westerkamp2024sn1006}, the reconstruction of SN1987A from eROSITA first-light data \citep{eberle2026bayesian}, and point-source extraction on the same dataset \citep{guardiani2025latent}. 
These works demonstrate the capability of \ac{IFT} for high-fidelity imaging and uncertainty-aware inference. 
The codebase developed in the course of these X-ray applications laid the groundwork for J-UBIK~\citep{eberle2024j}, a JAX-based implementation of the Universal Bayesian Imaging Kit \citep{ensslin2025universal}.
In this work, we combine \ac{IFT}-based deconvolution with an efficient, structured representation of spatially variant \acp{PSF}, thereby enabling spatially variant deconvolution of Chandra observations without introducing artifacts originating from noise overfitting. 
We demonstrate this through an application to real Chandra X-ray Observatory data of the supernova remnant Cassiopeia A.
Our approach not only improves image reconstruction quality but also provides principled uncertainty estimates, which are essential for scientific inference such as accurate surface brightness profiling. 

The structure of this work is as follows. We first introduce the problem of image deblurring (Sec.~\ref{sec.:blurring}) and discuss two approaches to its solution, namely \ac{IFT}-based (Sec.~\ref{sec.:ift}) and \ac{RL}-based methods (Secs.~\ref{sec.:RL} to \ref{sec.:RL_tik}). Furthermore, we discuss the dominant instrumental effects of Chandra (Sec.~\ref{sec.:Chandra}) and how we model them (Secs.~\ref{sec.:fast_psf} and \ref{sec.:PSFmodeling}).
We then benchmark \ac{IFT}-based Bayesian deconvolution on a synthetic example against three different \ac{RL} algorithms (Sec.~\ref{sec.:deconv_valid}).
Next, we denoise simulated Chandra \acp{PSF} (Sec.~\ref{sec.:fit_psf}) and investigate how well the chosen approach for approximating spatially varying \acp{PSF} can recover the complex structure of the Chandra \ac{PSF} (Sec.~\ref{sec.:psf_valid}). Finally, we apply the developed methods to Chandra data of Cassiopeia A (Sec.~\ref{sec.:ChandraApp}).

\section{Methods}
In this section we explain the problem of blurring and deblurring, as well as the methods used in this work to address it. 
\subsection{Blurring and Deblurring}\label{sec.:blurring}
Blurring effects such as optical aberrations or diffraction patterns degrade image quality across a wide range of scales, from microscopy and photography to astronomical imaging. 
Mathematically these effects can be expressed by an integral equation,
\begin{equation}\label{eq.:freqholmfull}
  d(y) = \int a(x,y) s(x) \ \text{d}x \ ,
\end{equation}
where $s(x)$ is the true object, $a(x,y)$ the blurring kernel, and $d(y)$ the observed blurred image.
From a probabilistic perspective, the blurring kernel $a(x, y)$, often called \ac{PSF}, is just the conditional probability $\mathcal{P}(y \mid x)$ of an incident photon at a position $x$ to be spread out or deflected to another position $y$. 
This means that this operation should be flux-conserving, provided no flux is lost at the boundaries,
\begin{equation}\label{eq.:psf_normalization}
\int_{\Omega_y} a(x, y) \ \mathrm{d}y = \int_{\Omega_y} \mathcal{P}(y \mid x) \ \mathrm{d}y = 1 .
\end{equation}
On a discrete grid, integral equation Eq.~\eqref{eq.:freqholmfull} turns into a matrix-vector product,
\begin{equation}\label{eq.:blurring}
    d = As \ .
\end{equation}
To retrieve the original $s$ from the blurred $d$, we need to invert $A$:
\begin{eqnarray}
    s = A^{-1}d \ .
\end{eqnarray}

If $a(x,y)$ can be assumed to depend only on the distance between two points, i.e., $a(x,y)\approx a(x-y)$, the Fredholm integral equation Eq.~\eqref{eq.:freqholmfull} can be approximated as a convolution\footnote{Here, $\ast$ denotes convolution.},
\begin{equation}\label{eq.:convolution}
  d(y) = \int a(x-y) s(x) \, \mathrm{d}x  = (a \ast s)(y)\, .
\end{equation}
Using the convolution theorem, the spatially invariant \acp{PSF} are just a multiplication in Fourier space, where $\mathcal{F}$ is the associated Fourier transformation,
\begin{equation}
  a \ast s  = \mathcal{F}^{-1}\left[ \ \mathcal{F}(a) \ \mathcal{F}(s) \ \right] \ .
\end{equation}
On a discrete grid $a(x-y)$ is a circulant matrix, which gets diagonalized by Fourier transformation.
This representation gets very efficient due to the efficiency of the fast Fourier transformation~\citep{cooley1965algorithm}.

Regardless of the spatial variability of \acp{PSF}, deblurring is an inherently ill-posed problem.
For $A$ to be invertible, all its eigenvalues must be non-zero.
Typically, this is not true for \acp{PSF}, which act as a low-pass filter, suppressing frequencies of small spatial scales. 
The suppression of high frequencies corresponds to a loss of information, which in the worst case renders $A$ singular.
However, even in the case where the \ac{PSF} does not completely erase certain high-frequency components, it substantially reduces power on these corresponding small scales, which increases the condition number of the convolution operator, defined as the ratio between its largest and smallest eigenvalue.
This large condition number makes numerical inversion inaccurate or even infeasible.
This is already a problem for noiseless data.

In practice, however, we are usually dealing with noisy data.
Due to the noise, there is no unique solution, but only a distribution of possible solutions that are statistically consistent with the observation.
Both of these effects, blurring and low-\ac{S/N}, render the problem ill-posed.
Nevertheless, incorporating prior knowledge can render this inversion can become feasible. 
For some algorithms this can be encoded through early stopping criteria, as discussed in Sec.~\ref{sec.:RL}, regularization terms, see Sec.~\ref{sec.:RL_tv}, \ref{sec.:RL_tik}, or directly with generative modeling, see Sec.~\ref{sec.:ift}. 
Other deblurring algorithms rely on neural networks, which require prior training on representative data, and thus implicitly encode a prior.

\subsection{Information Field Theory}\label{sec.:ift}
Information field theory~(\ac{IFT})~\citep{ensslin2009} is a Bayesian framework for inferring continuous physical fields, such as brightness distributions $s$, from discrete observational data $d$. In general, instruments do not directly probe these fields but instead access them indirectly through finite, discrete observables such as photon counts. 
Reconstructing a continuous field from such finite, discrete data leads to an ill-posed inverse problem.
The measurement process is modeled as
\begin{equation}\label{eq:measurement_easy}
  d = \mathcal{R}(s) + n\, ,
\end{equation}
where $\mathcal{R}$ denotes the measurement operator and $n$ represents noise. 
Using Bayes' theorem, the posterior distribution over the field is given by
\begin{equation}\label{eq:bayes}
  \mathcal{P}(s \mid d) = \frac{\mathcal{P}(d \mid s)\,\mathcal{P}(s)}{\mathcal{P}(d)} \, .
\end{equation}

In many inference problems, it is convenient to work with negative log-probabilities, which are called information Hamiltonians in \ac{IFT}. Analogous to Eq.~\eqref{eq:bayes}, the posterior information Hamiltonian $\mathcal{H}(s \mid d)$ reads as 
\begin{equation}\label{eq:hamiltonian_bayes}
  \mathcal{H}(s \mid d) = \mathcal{H}(d \mid s) + \mathcal{H}(s) - \mathcal{H}(d)\, .
\end{equation}
Taking the logarithm of the product rule of probabilities (see Eq.~\eqref{eq:bayes}) shows that information, whether from prior knowledge or from multiple measurements, combines additively, as reflected in Eq.~\eqref{eq:hamiltonian_bayes}.
This means that once a signal of interest is defined, it can in principle be observed multiple times or with different instruments. Each observation provides additional information about $s$ to update the posterior accordingly.

In the following we will use the Python package \ac{NIFTy} to build generative, standardized models.
By standardized we mean that the generative model includes a coordinate transformation $s(\xi)$, such that samples $\xi \sim \mathcal{N}(0,\mathbf{1})$ are mapped to a field $s$ whose distribution $\mathcal{P}(s)$ reflects our prior knowledge of the problem. 
This enables us to use inference algorithms like \ac{MGVI} by \cite{knollmuller2019metric}, which relies on the standardization of the prior.

The basic concept of variational inference is to minimize the Kullback-Leibler divergence,
\begin{align}
  \mathcal{D}_{\mathrm{KL}} \left(\mathcal{Q}_{\bar{\xi},\,  \Xi}(\xi) \parallel  \mathcal{P} (\xi \mid d)\right) = \int \mathcal{D}\xi \, \mathcal{Q}_{\bar{\xi},\, \Xi}(\xi) \ln\frac{\mathcal{Q}_{\bar{\xi},\,  \Xi}(\xi)}{\mathcal{P}(\xi \mid d)} \, ,
\end{align}
between the true posterior and a family of distributions. In Gaussian variational inference this family of distributions consists of normal distributions $\mathcal{Q}_{\bar{\xi},\, \Xi}$, with mean $\bar{\xi}$ and covariance $\Xi$. Since $\Xi$ scales quadratically with the number of model parameters, in MGVI we use an implicit operator representation of the inverse Fisher information metric as an approximation of the covariance.
This enables the algorithm to take correlations between all model parameters into account, while still scaling linearly.
The integral is then estimated via sampling.
In MGVI, we use antithetic sampling, i.e., for each drawn sample $\xi_i \sim \mathcal{N}(\bar{\xi},\Xi)$, we additionally include its reflection about the posterior mean, $\xi_i^{\, -} \coloneqq 2\bar{\xi} - \xi_i$.
Since the approximate posterior is Gaussian and thus symmetric around its mean, this ensures that the sample mean exactly matches $\bar{\xi}$, thereby reducing the variance of the stochastic estimate of the KL-divergence and its gradient.
The result of MGVI is a Normal distribution $\mathcal{Q}_{\bar{\xi}, \Xi}$, defined by the posterior mean $\bar{\xi}$ and the local metric $\Xi$, from which we draw $N_s$ posterior samples.

These samples are used to represent the posterior. 
For any observable $f$ in the forward model, we can estimate its posterior mean $m(f)$ and standard deviation $\sigma(f)$ via
\begin{align}
  m(f) &= \langle f \rangle_{\mathcal{P}(\xi \mid d)} \approx \frac{1}{N_s} \sum_i f(\xi_i) \,, \\
  \sigma(f) &= \sqrt{\left\langle \left(f - m(f)\right)^2 \right\rangle_{\mathcal{P}(\xi \mid d)}} \approx \sqrt{\frac{1}{N_s} \sum_i \left(f(\xi_i)-m(f)\right)^2} \,.
\end{align}
Here, $f$ can be any function defined on the latent space, allowing us to evaluate posterior statistics at any stage of the forward model, e.g., the mean and standard deviation of the signal $s$ itself, or of the signal response $\mathcal{R}(s)$. The latter can be directly compared to the data.

\subsubsection{Poisson likelihood}\label{sec.:likelihood}
In the specific case of photon-counting experiments such as X-ray observations, we do not use the simplified model of Eq.~\eqref{eq:measurement_easy}, but instead specify the likelihood explicitly as the Poisson distribution. 
For multiple detector pixels $i$, the counts are independent, so that the joint likelihood is the product of pixel-wise Poisson distributions,
\begin{equation}\label{eq.:measurement_poisson}
\mathcal{P}(d \mid \lambda) = \prod_i \frac{\lambda_i^{d_i} e^{-\lambda_i}}{d_i!} \quad \text{with} \quad \lambda_i = \mathcal{R}_i(s) \, .
\end{equation}
The map $\mathcal{R}$ from $s$ to $\lambda$ describes all instrumental effects, while the Poisson likelihood accounts for the stochastic nature of the measurement process.
This stochastic effect, known as Poisson noise or shot noise, arises because photon detection is an inherently discrete, random process, even when the expected count rate
$\lambda$ is known exactly.
Throughout this work, we use the word noise solely for this measurement uncertainty, rather than for unwanted additional signals.
The obtain the likelihood Hamiltonian for Eq.~\eqref{eq:hamiltonian_bayes}, we compute the \ac{NLL},
\begin{equation}\label{eq.:hamiltonian_likelihood}
  \mathcal{H}(d \mid s) = -\log \mathcal{P}(d \mid s)
  = \sum_i \left( \lambda_i - d_i \log \lambda_i + \log d_i! \, \right).
\end{equation}
The derived likelihood Hamiltonian is highly related to the often used Cash statistics~\citep{cash1979parameter}.
If the only instrumental effect is blurring, such that $\mathcal{R}(s)=a \ast s $, Eq.~\eqref{eq.:hamiltonian_likelihood} reads as:
\begin{align}\label{eq.:hamiltonian_likelihood_simple}
  \mathcal{H}(d \mid s) 
  &= \sum_i (a\ast s)_i - d_i \log (a\ast s)_i + \log d_i! \,.
\end{align}
However, in reality the map $\mathcal{R}$ from the true field $s$ to the signal response $\lambda$ consists of numerous instrumental effects, such as spatially variant \ac{PSF}, effective area, and detector response. 
A detailed description of these effects for the Chandra telescope is given in Sec.~\ref{sec.:Chandra}. 

Extending Eq.~\eqref{eq.:measurement_poisson} and Eq.~\eqref{eq.:hamiltonian_likelihood_simple} to multiple, independent measurements of the same signal $s$ with different instrumental effects yields
\begin{align}\label{eq.:hamiltonian_likelihood_more}
  \mathcal{H}(d \mid s) &= \sum_j \mathcal{H}(d_j \mid s) = \sum_j -\log \mathcal{P}(d_j \mid s) \\
  &= \sum_j \sum_i \left( (a_j\ast s)_{i} - d_{ij} \log (a_j\ast s)_{i} + \log d_{ij}! \right) \,.
\end{align}
This enables us to do inference with multiple data sets and potentially differing instrumental descriptions, here $a_j$.

\subsubsection{Correlated field prior}
To perform inference, prior knowledge about $s$ must be specified via the prior distribution $\mathcal{P}(s)$. In the present case, we aim to encode only minimal physical assumptions. First, we require the field to vary smoothly in space, i.e., to exhibit spatial correlations. Second, the field $s$ is constrained to be strictly positive, as it represents a physical flux.

\begin{figure}
  \includegraphics{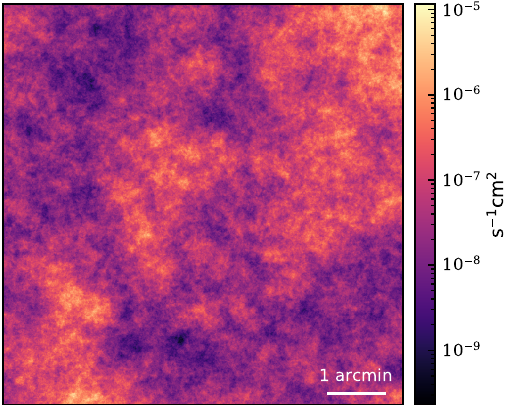}
  \caption{Prior sample of the log-normal correlated photon flux field model, see Eq.~\eqref{eq.:log_normal_field}.}\label{fig.:prior_sample}
\end{figure}

The first constraint can be fulfilled by modeling a correlated normal field,
\begin{equation}\label{eq:correlated_gaussian}
  \phi \sim \mathcal{N}(0, \Phi) \, ,
\end{equation}
with covariance $\Phi$.
Rather than specifying this prior directly, we construct a hierarchical generative model in a latent space formulation. We introduce a latent variable $\xi$, drawn from a standard normal distribution,
\begin{equation}
  \xi \sim \mathcal{N}(0, \mathbf{1}) \, .
\end{equation}
The covariance $\Phi$ can always be factorized as $\Phi = \mathcal{AA}^\dagger$, and thus samples from Eq.~\eqref{eq:correlated_gaussian} can be generated via the linear transformation
\begin{equation}\label{eq:correlated_transform}
  \phi = \mathcal{A} \, \xi \, .
\end{equation}
In general, this operation scales quadratically with the number of degrees of freedom. However, under the assumption of statistical homogeneity, the covariance $\Phi$ is diagonal in Fourier space (Wiener--Khinchin theorem) and can be parametrized by the power spectrum $P(k)$ via
\begin{equation}
  \Phi = \mathcal{F}^{-1} \, \mathrm{diag}\left(P(k)\right) \, \mathcal{F} \ .
\end{equation}
This reduces the number of degrees of freedom and allows for an efficient parametrization in terms of the power spectrum $P(k)$, or equivalently its square root, the amplitude spectrum $\sqrt{P(k)}$. The amplitude spectrum $\sqrt{P(k)}$ relates to $\mathcal{A}$ as
\begin{equation}
  \mathcal{A} = \mathcal{F}^{-1}\mathrm{diag}\left(\sqrt{P(k)}\right) \mathcal{F} = \mathcal{F}^{-1}\hat{\mathcal{A}}\mathcal{F}\, ,
\end{equation}
where $\hat{\mathcal{A}} = \mathrm{diag}\left(\sqrt{P(k)}\right)$ is the representation of $\mathcal{A}$ in Fourier space.
Using this, Eq.~\eqref{eq:correlated_transform} becomes
\begin{equation}\label{eq:correlated_transform_harmonic}
  \phi = \mathcal{F}^{-1}\bigl( \hat{\mathcal{A}} \, \hat{\xi} \bigr) \, , \quad \text{with} \quad \hat{\xi} = \mathcal{F}\xi \, .
\end{equation}
The amplitude spectrum is itself parameterized in order to encode physical prior information, such as an underlying power-law behavior with flexible deviations. In a hierarchical formulation, $\hat{\mathcal{A}}$ can be modeled as a function of additional latent parameters $\xi_\mathcal{A}$, allowing for adaptive inference of the power spectrum within the same framework (see \citep{arras2022variable} for details).
To enforce positivity, the physical field is obtained via a point-wise nonlinear transformation, e.g.,
\begin{equation}\label{eq.:log_normal_field}
  s = s_0 \exp(\phi) \,,
\end{equation}
where $s_0$ is spatially constant and carries the units of the field $s$. 
This defines a flexible hierarchical model that represents a broad class of correlated random fields with an inferred power spectrum, combined with a positivity constraint. 
Fig.~\ref{fig.:prior_sample} shows a prior sample of the log-normal correlated field. 

\subsection{Richardson-Lucy deconvolution}\label{sec.:RL}
One of the most widely used algorithms for deconvolution was independently proposed by \citet{richardson1972bayesian} and \citet{lucy1974iterative}, and is therefore known as the \ac{RL} deconvolution. 
The algorithm describes an iterative scheme to recover a deconvolved image from the observed data $d$ and the \ac{PSF} $a$, starting from an initial guess $s^{(0)}$:
\begin{equation}\label{RLdeconv}
s^{(n+1)} = s^{(n)} \cdot \left[ \frac{d}{s^{(n)} \ast a} \ast \tilde{a} \right] = s^{(n)} \cdot c^{(n)} \,.
\end{equation}
Here, $\cdot$ denotes a pointwise product, $c^{(n)}$ is the correction term applied to the previous iterate, and $\tilde{a}$ denotes the flipped PSF, defined by $\tilde{a}_{j-i} = a_{i-j}$.
Derivations can be found in App.~\ref{app.:regularization} and \citep{lucy1974iterative}.

The algorithm guarantees positivity of the solution for $s^{(n)} \ge 0$ and, under the assumption of Poisson noise, converges to a maximum-likelihood estimate. 
In the absence of explicit regularization, however, the method is prone to overfitting and the number of iterations acts as an implicit regularization parameter. 
The optimal number of iterations depends on the noise level: for data with high-\ac{S/N}, more iterations can be applied, whereas for noisy data, early stopping is required to prevent noise amplification.
Too many iterations introduce artifacts, which typically appear first in low-\ac{S/N} regions, while too few iterations limit the recovery of the underlying signal.
This limitation becomes particularly severe for data with a high dynamic range, where the optimal number of iterations varies across the image.

\subsection{Regularization and prior knowledge}
The RL algorithm can be extended by an additional regularization term $\mathcal{J}$. 
From a Bayesian perspective, this regularization acts like a prior in the iterative scheme. 
By minimizing the posterior Hamiltonian, we can access a \ac{MAP} solution.
Since the evidence term is independent of the free parameters, we can minimize the joint Hamiltonian instead,
\begin{align}\label{eq.:regularization}
  \mathcal{H}(s,d) &= \mathcal{H}(d \mid s)  + \mathcal{H}(s)\\
  &= \mathcal{H}(d \mid s) + \mu \ \mathcal{J} (s) \, .
\end{align}
The importance of the prior is controlled by $\mu$, which is typically set manually or via heuristics.
A variety of regularization terms have been proposed for RL-based imaging. In this work, we use total variation (see Sec.~\ref{sec.:RL_tv}) and Tikhonov regularization (see Sec.~\ref{sec.:RL_tik}).

\subsection{Total variation regularized Richardson-Lucy}\label{sec.:RL_tv}
The \ac{TV} regularization, first considered by~\cite{rudin1992nonlinear} for denoising purposes, was applied to deblur images by~\cite{dey_richardsonlucy_2006}.
The \ac{TV} regularization extends the RL algorithm by a factor $r^{(n)}$ which punishes the gradient of $s^{(n)}$ and therefore smoothens the result,
\begin{equation}\label{eq.:RL_tv}
s^{(n+1)} = s^{(n)} \cdot c^{(n)} \cdot \left[1 - \mu \, \mathrm{div} \, \left(\frac{\nabla s^{(n)}}{\lvert\nabla s^{(n)}\rvert}\right)\right]^{-1}\ = s^{(n)} \cdot \frac{c^{(n)}}{r^{(n)}} \ .
\end{equation}
One can show that this iterative scheme returns a \ac{MAP} result for Eq.~\eqref{eq.:regularization} (see App.~\ref{app.:regularization}) with a $\mathcal{J}(s)$ of the form:
\begin{equation}
  \mathcal{J}_\mathrm{TV}(s) = \lVert\nabla s\rVert_{2,1} \,.
\end{equation}
The mixed gradient norm\footnote{The $(2,1)$-norm is calculated by first taking the $2$-norm of the two-dimensional gradient vector at each pixel, and then the 1-norm of the resulting magnitudes over all pixels.} makes sure that the penalty grows for solutions that are non-smooth, but also enforces sparsity. This means that the resulting image is piece-wise smooth, but can still have some sharp features.

\subsection{Tikhonov regularized Richardson-Lucy}\label{sec.:RL_tik}
The second regularization term that we use is a Tikhonov regularization of the form:
\begin{equation}
\mathcal{J}_{\mathrm{TIK}} = \lVert\nabla s \rVert^2_2 \,.
\end{equation}
This can be interpreted as a Gaussian prior with a $k^{-2}$ falling power spectrum.
Here, we penalize the gradient again, which smooths the solution. In contrast to TV regularization, the $\ell_2$-norm suppresses high gradient values, such as edges, more strongly, resulting in a smoother solution. The Tikhonov regularization can be implemented efficiently in Fourier space:
\begin{equation}\label{eq.:RL_tik}
s^{(n+1)} = \mathcal{F}^{-1}\left[ \left(1 + \mu \lvert k\rvert^2\right)^{-1}\mathcal{F}\left(s^{(n)} \cdot c^{(n)}\right)\right] \ .
\end{equation}
A detailed derivation is given in App.~\ref{app.:regularization}.

\subsection{Chandra instrument model}\label{sec.:Chandra}
The problem of imaging with X-ray observations poses additional challenges beyond deblurring, which we address by modeling the full instrument response.
The Chandra X-ray Observatory collects photons with a Wolter telescope and projects them onto a CCD detector. We describe the measurement process by a linear model $\mathcal{R}$,
\begin{equation}
  \lambda = \mathcal{R}(s) = \left(M \circ E \circ A \right)(s) \ ,
\end{equation}
where $\circ$ denotes the composition operator, $M$ is the mask, $E$ is the exposure, and $A$ is the \ac{PSF}. 
Each component is described below. 
Note that this model neglects non-linear effects such as pile-up.

The \ac{PSF} operator $A$ describes the blurring effect, as introduced in Sec.~\ref{sec.:blurring}.
This blurring is mainly caused by two factors. The first factor is the geometry of the Wolter telescope and the arrangement of the mirrors.
The degradation for Wolter telescopes mostly depends on the off-axis angle of the incident photons, but also on the azimuth angle. 
The larger the off-axis angle, the larger the \ac{PSF}, which leads to a more severe blurring of the incoming signal $s$.
This makes $A$ incompatible with the assumption of spatial invariance.
The second factor is that the charge released in the detector by an incoming X-ray photon is not always correctly reconstructed in the right pixel. 
This also leads to some bleeding into neighbouring pixels. 
The combination of these two effects determines the shape of the \ac{PSF}.
As described in Sec.~\ref{sec.:blurring}, the \ac{PSF} is often approximated by a convolution, assuming that it does not vary along the chosen \ac{FOV}. This assumption of spatial invariance of the \ac{PSF} only holds over a narrow part of the \ac{FOV}.
Providing reliable deblurring over the full \ac{FOV} is a primary goal of this work, and we explain our approach in Sec.~\ref{sec.:fast_psf}.

The second effect describes the sensitivity of the detector at a given position. Therefore, we define the exposure $E$. 
The longer we observe a certain object, the more photons we collect under the assumption of constant flux over time. 
Due to gaps between the detector panels, and flagged detector times, we will get an inhomogeneous exposure, which leads to inhomogeneous noise statistics even for areas with constant spatial flux.
In the forward model, these effects can be taken care of by a multiplication with effective observation time and sensitivity, here called $E$.

Additionally, some pixels can be defect. These pixels are flagged as defective in the Chandra \ac{CALDB}, and have zero exposure time $E$. If we recall the likelihood in Eq.~\eqref{eq.:hamiltonian_likelihood}, we see that a zero value of $\lambda$ would cause the logarithm to be undefined. Since there is no exploitable information for not measured pixels, we decided to flag these in the forward model and remove them from the likelihood calculation. To be consistent in the number of dimensions, the same pixels are flagged also in the data. This selection operation is done by the mask, $M$. 

Together, these operators capture the dominant instrumental effects that shape the observed data, allowing us to accurately model the relationship between the true sky signal
$s$ and the expected photon counts $\lambda$. 
Combined with the likelihood $\mathcal{P}(d \mid s)$ defined in Sec.~\ref{sec.:likelihood}, this enables the Bayesian inference of the signal $s$.

\subsection{Linear patched convolution}\label{sec.:fast_psf}
As discussed in Sec.~\ref{sec.:blurring}, a spatially invariant \ac{PSF} can be represented efficiently with convolution.
In the specific case of astrophysical imaging with X-ray observatories like Chandra, but also eROSITA or XMM-Newton, the property of spatial invariance of the \ac{PSF} only holds for quite small \acp{FOV}.
For larger fields of view, where this does not hold, we cannot assume the blurring to be a pure convolution.
In the most general case, the blurring operator $A$ would need to be represented as a full matrix. To illustrate why this is computationally infeasible, consider the blurring of a $1024 \times 1024$ pixel image with $\mathcal{O}(10^6)$ pixels: the corresponding full matrix would have $\mathcal{O}(10^{12})$ entries, requiring approximately 8 $\mathrm{TByte}$ of memory in a 64-bit floating point representation\footnote{$10^{12} \times 64\,\mathrm{Bit} \times \frac{1}{8}\,\mathrm{Byte}\,\mathrm{Bit}^{-1} \approx 8\,\mathrm{TByte}$.}.
Loading this matrix from disk alone would already exceed practical memory constraints. 
Furthermore, all discussed inversion techniques require multiple applications of the blurring operator, increasing the computational load further.
Specifically for the IFT approach, we require the blurring operation to be memory-efficient and fast.
Storing or applying the full matrix $A$ is therefore infeasible, yet we aim to improve upon the spatially invariant approximation.

To balance physical accuracy and computational efficiency, we approximate $A$ by $A^\star$, an interpolation of $N_k$ \ac{PSF} samples $c_k$ for specific positions in the field of view: 
\begin{equation}\label{eq.:matrix decomposition}
  A_{ij} \approx A^\star_{ij} = \sum_k^{N_k} w_{k,j} \, c_{k,i-j} \quad \mathrm{where} \quad \sum_k w_{k,j} = 1 \quad \forall j \ . 
\end{equation}
Here, $i$ and $j$ denote vectors to the detection and source pixel positions, and $c_{k,\, i-j}$ gives the probability of a photon emitted at $j$ being detected at $i$ under the $k$-th \ac{PSF} sample, while the weights
$w_{k,j}$ interpolate between these reference PSFs across the field of view.
The interpolation can be implemented in two ways: weighting the image and then convolving it, or convolving it first and then weighting it.
In this work, we adopt the first approach, since it is flux-conserving (see App.~\ref{app.:flux_cons} for details).
Without further simplifications, the computational cost of this approach scales linearly with the number of different \acp{PSF} $N_k$.
However, it is in principle capable of representing the true blurring operation $A$, since in the limit $N_k \rightarrow N_\mathrm{pix}$ each pixel has its own dedicated \ac{PSF} sample, thus recovering the full matrix.
The optimal choice of $N_k$ represents a trade-off between accuracy and computational cost.

A variety of weighting and \ac{PSF} selection schemes exist, each relying on different assumptions and simplifications.
To increase the accuracy of the \ac{PSF} representation of Chandra, we use the linear interpolation of \acp{PSF} in adjacent regions, as described in \citet{nagy1997fast}, since it provides a good compromise between computational effort and accuracy. 
We refer to this method as linear patched convolution.
The approach combines the matrix decomposition of Eq.~\eqref{eq.:matrix decomposition} with an overlap-add convolution scheme, which exploits the fact that for \acp{PSF} with limited support, meaning they are small compared to the \ac{FOV}, multiple small convolutions are computationally cheaper than one large convolution.
Concretely, the image $s$ is first decomposed into overlapping patches by operator $P$, weighted by $W$, convolved with the corresponding \ac{PSF} sample by operator $C$, and finally recombined by $G$.
The full approach yielding the blurred signal $s^\prime$ is then
\begin{equation}
  \label{eq:psfapprox}
 s^\prime \approx  A^\star s = G C W P s \ .
\end{equation}

A more detailed description of the individual operations follows. 
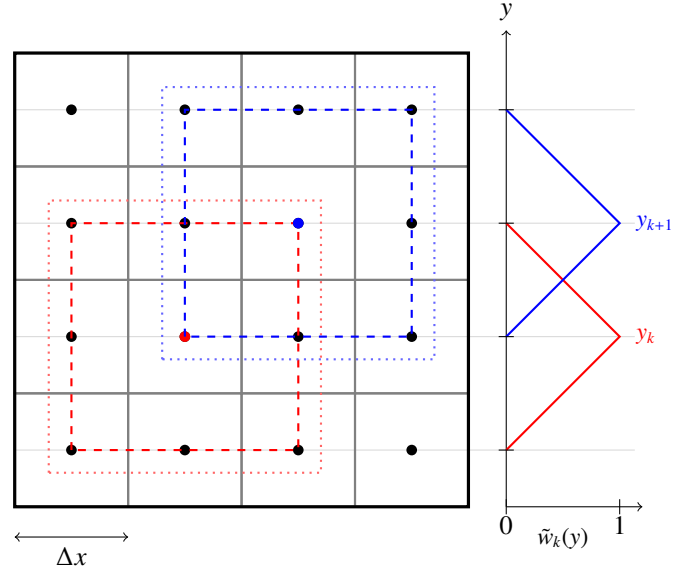
\begin{figure}
  \begin{tikzpicture}

\def\dx{1.5}  
\def\pad{0.3} 

\foreach \j in {0,1,2,3} {
    \draw[gray!30, thin] 
        (0, \j*\dx+0.5*\dx) -- (4*\dx+2.2, \j*\dx+0.5*\dx);
}

\foreach \i in {0,1,2,3} {
    \foreach \j in {0,1,2,3} {
        \draw[gray, thick] 
            (\i*\dx, \j*\dx) rectangle (\i*\dx+\dx, \j*\dx+\dx);
    }
}

\foreach \i in {0,1,2,3} {
    \foreach \j in {0,1,2,3} {
        \fill[black] (\i*\dx+0.5*\dx, \j*\dx+0.5*\dx) circle (2pt);
    }
}

\draw[red, thick, dashed] 
    ({1.5*\dx - \dx}, {1.5*\dx - \dx}) rectangle 
    ({1.5*\dx + \dx}, {1.5*\dx + \dx});
\fill[red] (1.5*\dx, 1.5*\dx) circle (2pt);

\draw[red!60, thick, dotted]
    ({1.5*\dx - \dx - \pad}, {1.5*\dx - \dx - \pad}) rectangle
    ({1.5*\dx + \dx + \pad}, {1.5*\dx + \dx + \pad});

\draw[blue, thick, dashed] 
    ({2.5*\dx - \dx}, {2.5*\dx - \dx}) rectangle 
    ({2.5*\dx + \dx}, {2.5*\dx + \dx});
\fill[blue] (2.5*\dx, 2.5*\dx) circle (2pt);

\draw[blue!60, thick, dotted]
    ({2.5*\dx - \dx - \pad}, {2.5*\dx - \dx - \pad}) rectangle
    ({2.5*\dx + \dx + \pad}, {2.5*\dx + \dx + \pad});

\draw[very thick] (0,0) rectangle (4*\dx, 4*\dx);

\draw[<->] (0, -0.4) -- (\dx, -0.4)
    node[midway, below] {$\Delta x$};

\def\ox{4*\dx + 0.5}  
\def\ph{1.5}          

\draw[->] (\ox, 0) -- (\ox, 4*\dx + 0.3) node[above] {$y$};
\draw[->] (\ox, 0) -- (\ox + \ph + 0.3, 0);

\node[below, font=\small] at (\ox + \ph/2, -0.15) {$\tilde{w}_k(y)$};

\draw[red, thick]
    (\ox,       0.5*\dx) --
    (\ox+\ph,   1.5*\dx) --
    (\ox,       2.5*\dx);

\draw[blue, thick]
    (\ox,       1.5*\dx) --
    (\ox+\ph,   2.5*\dx) --
    (\ox,       3.5*\dx);

\draw (\ox+\ph, -0.1) -- (\ox+\ph, 0.1) node[below=3pt] {$1$};

\draw (\ox, -0.1) -- (\ox, 0.1) node[below=3pt] {$0$};

\foreach \j in {0,1,2,3} {
    \draw (\ox-0.1, \j*\dx+0.5*\dx) -- (\ox+0.1, \j*\dx+0.5*\dx);
}

\node[red,  right, font=\small] at (\ox+\ph+0.1, 1.5*\dx) {$y_k$};
\node[blue, right, font=\small] at (\ox+\ph+0.1, 2.5*\dx) {$y_{k+1}$};

\end{tikzpicture}
  \caption{Illustration of the patching operator $P$ and the weighting operator $W$. The gray grid shows the natural tiling of the image into $N_\mathrm{patch}$ tiles. The dashed lines show the patch selected by $P$ for two different central points, highlighted in blue and red, respectively. The dotted lines depict the patch size after padding right before convolution with $c_k$. On the right, $\tilde{w}_k(y)$ is shown for two patches, which is one factor of the two-dimensional weighting function $w_k$, see Eq.~\eqref{eq.:weightingfunc}.}\label{fig.:linpatch}
\end{figure}
When we divide the \ac{FOV} into $N_\mathrm{patch}$ tiles, each tile has side length $\Delta x$ and $\Delta y$, where we assume $\Delta x= \Delta y$ for simplicity, see Fig.~\ref{fig.:linpatch}. 
Instead of using these tiles, we slice out a larger patch of side length $2\Delta x$, such that the borders of this patch are defined by the connection between the centers of the neighbouring tiles. 
This is illustrated by the dashed lines in Fig.~\ref{fig.:linpatch}. 
So $P_k s =s_k$ returns one of these larger patches per central point.
Patches at the border of the field of view are zero-padded to retain the same size. Every interior pixel appears in exactly four cut-outs.

The next operator $W_k s_k = w_k \cdot s_k $ applies a bilinear weighting to each patch via point-wise multiplication with $w_k(x,y)$, defined as 
\begin{equation}\label{eq.:weightingfunc}
  w_k(x,y) = \tilde{w}_k(y) \cdot \tilde{w}_k(x) = \Lambda\, \left(\frac{y-y_k}{\Delta y}\right) \cdot \Lambda\, \left(\frac{x-x_k}{\Delta x}\right)
\end{equation}
where $\Lambda(x)= \mathrm{max}(1-\lvert x \rvert,0)$ denotes the triangular function, see Fig.~\ref{fig.:linpatch}.
Consequently, each pixel is weighted according to its distance to the patch center.

In the next step, we convolve each of these patches with its associated \ac{PSF}, $c_k$.
Since we use the convolution theorem, we need to correct for the implied but unphysical periodic boundary conditions.
To break these boundary conditions, we pad each patch with zeros, obtaining an additional margin (see dotted line in Fig.~\ref{fig.:linpatch}).

$G$ is a glue operator that stitches together the patches separated by $P$. 
Due to the additional padding before convolution, the operator $G$ is not simply the adjoint of $P$.
Since the patches were enlarged, the gluing operation also takes these additional pixels into account.
In the case that all $c_k$ are the same \ac{PSF}, this reduces to an overlap-add algorithm for convolution, which is commonly used when the support of one of the convolution terms is much smaller than the other.

\subsection{\texttt{MARX}-simulation and modeling of the PSF}\label{sec.:PSFmodeling}
To use the representation of the spatially variant \ac{PSF} discussed in Sec.~\ref{sec.:fast_psf}, we need the \ac{PSF} at each patch center. 
We use the program \texttt{MARX}~\citep{davis2012raytracing}\footnote{\url{https://chandra-marx.github.io/}} to simulate the Chandra \ac{PSF}, obtaining estimates $\tilde{c}_k$. 
Details on the MARX simulation setup are given in App.~\ref{app.:marx}.
The simulations produce event files containing a user-specified number of photon detection events, each with associated coordinates.
After binning these events on a grid, the simulated \acp{PSF} follow a per-pixel Poisson distribution and therefore suffer from shot noise in the low-flux regions.
\begin{figure}[h]
  \includegraphics{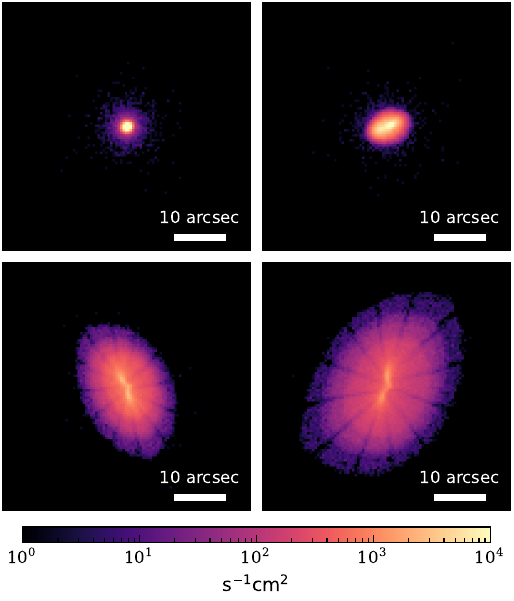}
  \caption{Examples of \texttt{MARX}-simulated \acp{PSF} across the \ac{FOV}. The \ac{PSF} changes from compact and circular at the optical axis (top-left) to elongated and structured at the edge (bottom-right). In all panels, the Poisson noise is clearly visible.
  }\label{fig.:psfs_raw}
\end{figure}

To obtain shot-noise-free \acp{PSF} from the \texttt{MARX}-simulations, we model each of them with a log-normal field (see Eq.~\eqref{eq.:log_normal_field}), 
\begin{equation}\label{eq:psf_model}
  \gamma = \exp\left(\mathcal{F}^{-1}\bigl( \hat{\mathcal{A}} \, \hat{\xi} \bigr) \right)\ ,
\end{equation}
and use MGVI to compute the posterior mean $\langle \gamma\rangle_{\mathcal{P}(\gamma \mid \tilde{c})}$, yielding denoised versions of the simulated \ac{PSF} count map $\tilde{c}$.
The normalized and denoised \acp{PSF} $c_k$ are then defined as
\begin{equation}
 c_k(y) = \frac{\langle \gamma_k\rangle(y)}{\int \mathrm{d}y^\prime \, \langle \gamma_k\rangle(y^\prime)}  \, , \quad \text{such that} \quad \int \mathrm{d}y \, c_k(y) = 1 \, .
\end{equation}

Since MGVI starts from a random latent position, not every inference converges to a valid posterior, meaning that not every run fits the data equally well. Therefore, we extended the algorithm from a manual convergence check to an automatic one, which restarts from a new random latent position if the inference does not fulfill a predefined criterion. To define this criterion, we consider the mean of the noise-weighted squared residuals,
\begin{equation}
  \chi_\mathrm{norm}^2=\frac{1}{N_\mathrm{data}} \sum_i \frac{\left(d-\lambda\right)_i^2}{\sigma_{\lambda_i}^2} = \frac{1}{N_\mathrm{data}} \sum_i \frac{\left(d-\lambda\right)_i^2}{\lambda_i} \,,
\end{equation}
where $N_\mathrm{data}$ is the number of data points and, in the case of Poisson statistics, $\sigma_\lambda=\sqrt{\lambda}$, which gives the expression after the second equality sign. Our stopping criterion is $\chi_\mathrm{norm}^2<1.0$.

We use this criterion primarily as a sanity check to ensure the inference has moved away from its random initialization, rather than as a strict goodness-of-fit target. 
A reconstruction stuck near initialization is not yet informed by the data and typically fails to reproduce the structure of the data, resulting in $\chi^2_\mathrm{norm}$ values well above unity. If the criterion is not met, the reconstruction is discarded and restarted from a new random latent position.

We note that for the PSF datasets, $\chi^2_\mathrm{norm}$ is systematically below unity due to the large fraction of low-count pixels, for which the Poisson residuals are right-skewed, rather than approximately Gaussian. This does not indicate overfitting. Visual inspection of the residuals for several PSF reconstructions shows no unresolved structure, and the use of posterior samples together with resampling in MGVI structurally discourages fitting to individual noise realizations.

\section{Validation}
In this section, we validate the deconvolution capability of \ac{NIFTy} and assess the accuracy of the \ac{PSF} fitting and interpolation method. 
The evaluation is based on synthetic data, enabling controlled tests under known conditions. We compare \ac{NIFTy} with standard deconvolution algorithms, namely plain \ac{RL}, \ac{RL} with \ac{TV} regularization, and \ac{RL} with Tikhonov regularization, across two test cases with different \ac{S/N}.
The results are quantified using multiple metrics: \ac{RMSE}, \ac{SSIM}, and \ac{NLL}, providing a comprehensive assessment of reconstruction quality.

\subsection{Deconvolution validation}\label{sec.:deconv_valid}

\ac{NIFTy} has already been applied in several prior studies for image reconstruction in X-ray and gamma-ray astronomy \citep{eberle2026bayesian, westerkamp2024sn1006, guardiani2025latent,scheel2023multicomponent}.
However, a direct comparison with standard deconvolution methods, particularly in the context of noisy, low-\ac{S/N} data, is essential to establish its relative performance. 
While \ac{NIFTy} is capable of more than image deblurring, such as source decomposition and modeling of complex instrumental effects, this section focuses on its deconvolution performance.
We compare \ac{NIFTy} with three variants of the RL algorithm (Secs.~\ref{sec.:RL},~\ref{sec.:RL_tv},~\ref{sec.:RL_tik}). 
In addition to visual inspection, we quantify performance using the following three complementary metrics.

The \ac{RMSE} measures the pixel-wise deviation between the reconstructed image $s_\mathrm{est}$ and the ground truth $s_\mathrm{gt}$:
\begin{equation}
  \mathrm{RMSE}(s_\mathrm{gt}, s_\mathrm{est}) = \sqrt{\frac{1}{N_\mathrm{pix}}\sum \left(s_\mathrm{est}-s_\mathrm{gt}\right)^2}.
\end{equation}
A lower \ac{RMSE} indicates better agreement with the true source morphology.

The \ac{SSIM}~\cite{wang2004image} evaluates image similarity in three perceptually relevant aspects: luminance, contrast, and structural information. It is computed on $7\times7$ patches of the ground truth and reconstructed image, denoted $o$ and $u$, respectively. The patch means $\mu_o$, $\mu_u$, variances $\sigma_o^2$, $\sigma_u^2$, and covariance $\sigma_{ou}$ are used to compute the \ac{SSIM} as a product of three components:
\begin{align}
l(o,u) &= \frac{2\mu_o\mu_u + z_1}{\mu_o^2 + \mu_u^2 + z_1} \ , \quad
  c(o,u) = \frac{2\sigma_o\sigma_u + z_2}{\sigma_o^2 + \sigma_u^2 + z_2} \ , \\
  s(o,u) &= \frac{2\sigma_{ou} + z_3}{\sigma_o^2 + \sigma_u^2 + z_3} \ ,
\end{align}
where $z_1 = \left(d_1L\right)^2$ with $d_1 = 0.01$, $z_2 = \left(d_2L\right)^2$ with $d_2 = 0.03$, and $L$ is the dynamic range. The overall \ac{SSIM} is then given by:
\begin{equation}
  \mathrm{SSIM}(a, b) = \left[l\left(a,b\right)\right]^\alpha \cdot \left[c\left(a,b\right)\right]^\beta \cdot \left[s\left(a,b\right)\right]^\gamma,
\end{equation}
which reduces to the standard form:
\begin{equation}\label{eq.:ssim}
  \mathrm{SSIM}(o, u) = \frac{\left(2\mu_o\mu_u + z_1\right)\left(2\sigma_{ou} + z_2\right)}{\left(\mu_o^2 + \mu_u^2 + z_1\right)\left(\sigma_o^2 + \sigma_u^2 + z_2\right)},
\end{equation}
when $\alpha = \beta = \gamma = 1$ and $z_3 = z_2/2$.
This decomposition allows for a detailed assessment of reconstruction quality: the luminance term $l(a,b)$ quantifies the similarity of mean intensities, the contrast term $c(a,b)$ captures the similarity of intensity variations, and the structure term $s(a,b)$ reflects the preservation of spatial patterns such as edges and textures.
Together, these components provide a comprehensive evaluation of how well the reconstruction matches the ground truth in terms of perceptual quality, going beyond pixel-wise differences to assess the fidelity of morphological features.
Unlike pixel-wise metrics such as \ac{RMSE}, \ac{SSIM} accounts for human visual perception and is more sensitive to structural fidelity, making it particularly suitable for evaluating deconvolution algorithms in X-ray astronomy, where the recovery of fine-scale morphology is critical.

The \ac{NLL}, which corresponds to the likelihood term of the Hamiltonian (Eq.~\eqref{eq.:hamiltonian_likelihood}), quantifies how well the model fits the observed data. A lower NLL indicates a better fit. For optimal performance, \ac{RMSE} and NLL should be minimized, while \ac{SSIM} should be maximized.

\begin{figure*}
\centering
  \includegraphics[width=0.95\textwidth]{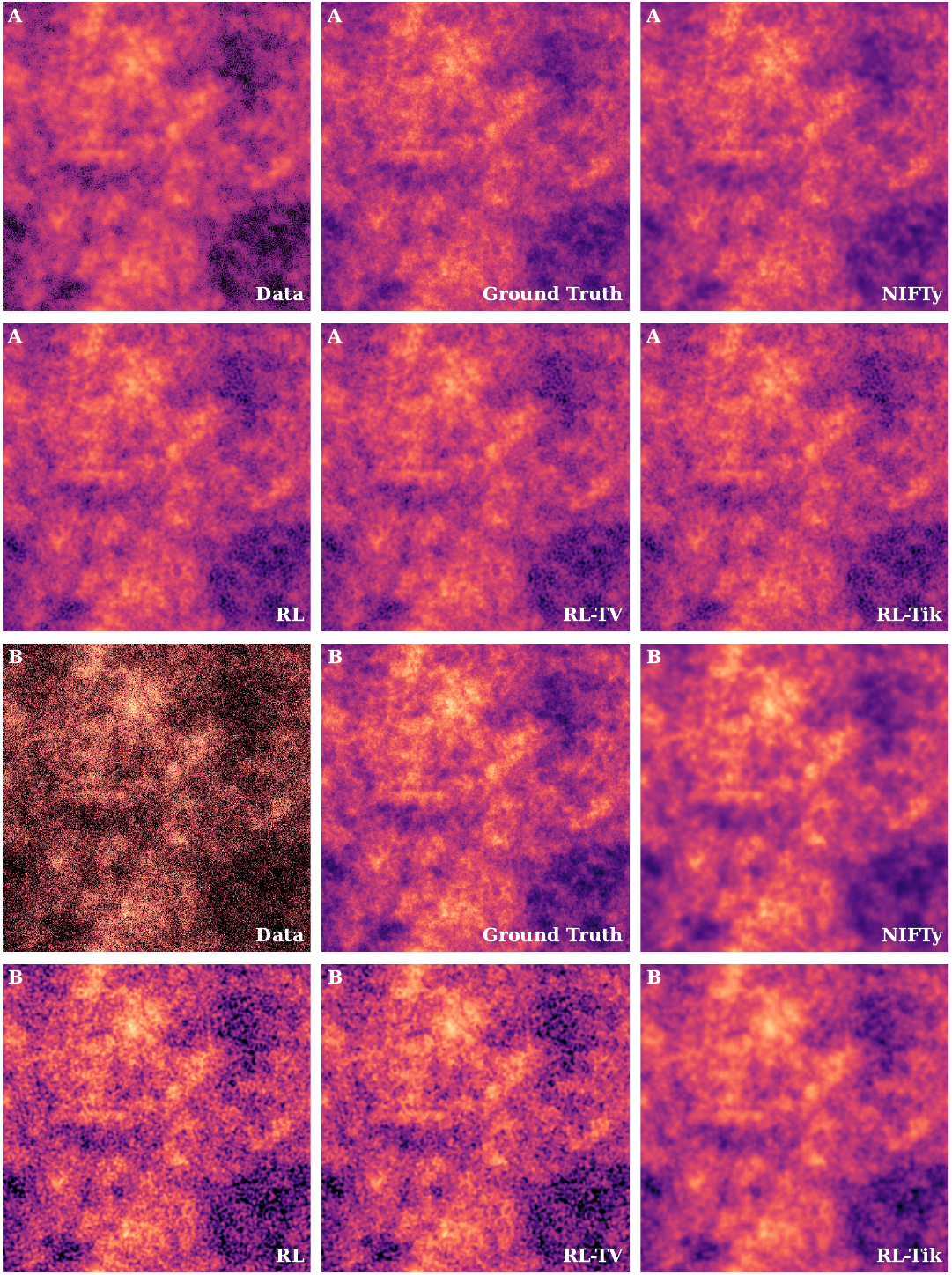}
   \caption{Comparison of deblurring methods. Labels indicate the hight and low \ac{S/N} cases (A and B), and distinguish between observed data, ground truth, and reconstruction results. Further metrics are provided in Table~\ref{tab.:deconv_comparison}.}\label{fig.:comparison_methods}
\end{figure*}

The test data consist of two synthetic images, A and B, generated from a log-normal field prior (Eq.~\eqref{eq.:log_normal_field}) using the same random seed, ensuring identical morphology. 
The images differ in dynamic range, resulting in different \ac{S/N} distributions (see Fig.~\ref{fig.:snrmaps}). 
They were convolved with an isotropic Gaussian \ac{PSF} with standard deviation $\sigma = 2$ pixels (FWHM $\approx 4.7$ pixels). 
The resulting expected count rate $\lambda$ is then used to draw Poisson-distributed data.
These noisy data serve as input for all reconstruction algorithms.
The reconstruction results are shown in Fig.~\ref{fig.:comparison_methods}, and quantitative performance metrics are summarized in Table~\ref{tab.:deconv_comparison}.

\begin{table}[ht]
\caption{Comparison of deconvolution methods for two different images A and B, see Fig.~\ref{fig.:comparison_methods}. Next to the \ac{RMSE} and \ac{SSIM} we also show the \ac{NLL}. To spot potential overfitting we added the NLL value between the blurred ground truth and the data as a reference value,"Ref".}
\label{tab.:deconv_comparison}
\begin{tabular}{llllr}
\toprule
 Image & Method & RMSE & SSIM & NLL \\
\midrule
\multirow[t]{4}{*}{A} & NIFTy & 4.3130 & 0.9330 & -5\,529\,800 \\
 & RL & 4.3878 & 0.9306 & -5\,533\,627 \\
 & RL Tik & 4.3877 & 0.9306 & -5\,533\,574 \\
 & RL TV & 4.3878 & 0.9306 & -5\,533\,625 \\
\midrule
 & Ref & & & -5\,524\,641 \\
\midrule
\midrule
\multirow[t]{4}{*}{B} & NIFTy & 0.1940 & 0.7181 & 234\,729 \\
 & RL & 0.2803 & 0.5538 & 226\,901 \\
 & RL Tik & 0.1973 & 0.7157 & 231\,152 \\
 & RL TV & 0.2568 & 0.5923 & 227\,810 \\
 \midrule
 & Ref & & & 233\,083 \\
\cline{1-5}
\bottomrule
\end{tabular}
\end{table}

\subsubsection{Performance at high S/N}
In Fig.~\ref{fig.:comparison_methods}, one can see that the ground truth of image A was blurred by the Gaussian \ac{PSF} and is severely affected by Poisson noise. 
Looking at the results from the different reconstruction methods, we can see that in the high-flux regions the results for all of the algorithms are in good agreement with the ground truth. 
In the low-flux regions however, we can clearly see that all RL-based methods leave unwanted artifacts due to noise amplification. 
RL-algorithms need to be stopped after a number of iterations, which depends on the noise level. 
In this case the algorithms were stopped after $n=9$ steps, at the point where the \ac{RMSE} was optimal. 
Specifically, this only means that we stopped at a point were the overall result was the best possible. In a real situation, this stopping criterion is not accessable.
For the higher-flux regions, we could have run it longer, while for the low-flux regions, we had already passed the optimal stopping point, leading to overfitting. Also $\lambda_\mathrm{TV}=10^{-5}$ and $\lambda_\mathrm{Tik}=3\cdot 10^{-3}$, were tuned such that we optimized for the best \ac{RMSE}. The effect of the regularization is barely visible.
Here, we see the advantage of \ac{NIFTy}, which is less likely to produce these artifacts. 
The way MGVI is designed makes sure that it is very unlikely to overfit a signal, since we optimize for an ensemble of samples, and draw new samples after each optimization step.
Therefore, we do not optimize for specific realizations.
From a theoretical point of view, this is due to the fact that we do not fit for a maximum posterior, but solve for an approximate posterior.
Looking at Tab.~\ref{tab.:deconv_comparison} we can state that the \ac{RMSE} and the \ac{SSIM} is best for \ac{NIFTy}.
This means that for two different criteria based on the deconvolution result, we get the best score with \ac{NIFTy}.
However, we get a worse value for the \ac{NLL} than the other methods. 
The \ac{NLL} describes the fit of the data. Here, the best outcome is the Ref-value of the \ac{NLL}, which is the \ac{NLL} produced by the ground truth. 
Surpassing this value describes an overfit. 
All methods exhibit slight overfitting, with \ac{NIFTy} being the least affected. Missing the optimal stopping point for \ac{RL} algorithms makes this overfitting worse and leads to more artifacts.
All in all, the metrics are quite similar, and we see that NIFTy is competitive with the standard algorithms.

\subsubsection{Performance at low S/N}
In Image B we have even more noise than in image A. 
For the RL Algorithm we found that the optimal point of stopping would be one step. 
Since the ground truth is unknown in practice, a user would not recognize this early overfitting and would realistically let the algorithm run longer. To reflect this, we let it run slightly longer, up to $n=3$.
We observe drastic overfitting of the noise structures.
This is supported by a lower NLL value than the ground truth, see Tab.~\ref{tab.:deconv_comparison}. 
The damage of overfitting could be reduced with the \ac{TV} regularization, $\lambda_\mathrm{TV}=0.07$. 
In the Tikhonov regularization, with $\lambda_\mathrm{Tik}=5$ we could improve the \ac{RMSE} even more. 
All of this however, either needs a lot of experience to prevent even more overfitting or knowledge about the ground truth.
With \ac{NIFTy} we could again, without the curse of overfitting, reconstruct a solution with less artifacts, which also outperforms the other algorithms with respect to \ac{RMSE} and \ac{SSIM}. 
Additionally, \ac{NIFTy} enables us to access the approximate posterior and thus gain additional insight into the space of possible solutions. In particular, we can sample from this posterior, look at different realizations, and calculate their mean and standard deviation.
Here, we only consider the posterior mean.

\subsubsection{Power spectra and histograms}
More illustrations regarding the quality of the inference algorithms can be found in App.~\ref{app.:deconv_metrics}. 
In the histogram and violin plots of Figs.~\ref{fig.:ImageAHist} and~\ref{fig.:ImageBHist}, we see that the error distribution for \ac{NIFTy} is less biased towards positive values and less spread out, especially in the noisy case of image B.
We further performed a power spectrum analysis, which shows the mathematically expected behavior. 
After blurring we lose power on short scales. By drawing from the Poisson distribution we get Poisson noise, which adds a lot of power on very short scales again. With RL we can recover parts of the lost power but we can recover even more by using \ac{NIFTy}.
\ac{RL} and \ac{RL}-\ac{TV} exhibit excess power at intermediate spatial scales, which manifests as ringing artifacts due to overfitting of the noise.

\subsection{Validation of the PSF denoising}\label{sec.:fit_psf}
After the validation of the deblurring capabilities of NIFTy compared to \ac{RL} algorithms, we want validate the representation of the Chandra \ac{PSF}. Therefore, we will first evaluate the denoising of the \texttt{MARX}-simulated \acp{PSF} $\tilde{c}_k$, yielding $c_k$.

\begin{figure}[h]
  \includegraphics{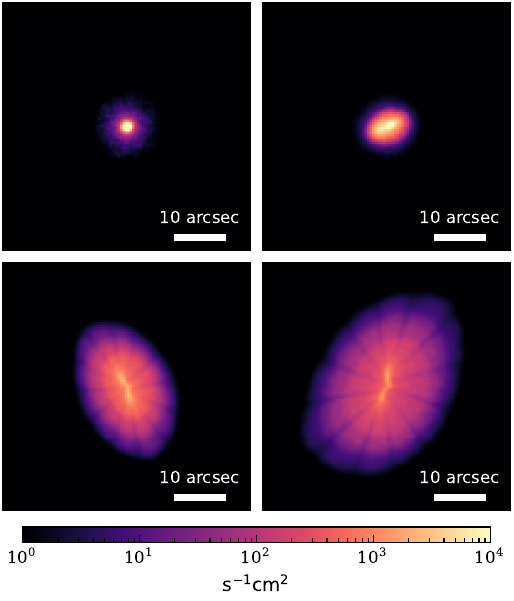}
  \caption{Denoised \acp{PSF} $c_k$. Compared to the \texttt{MARX}-simulated noisy data in Fig.~\ref{fig.:psfs_raw}, the model suppresses shot noise and accurately reproduces the observed substructures, demonstrating its capability to capture fine-scale morphology.}\label{fig.:psfs_fit}
\end{figure}
\begin{figure}
  \includegraphics{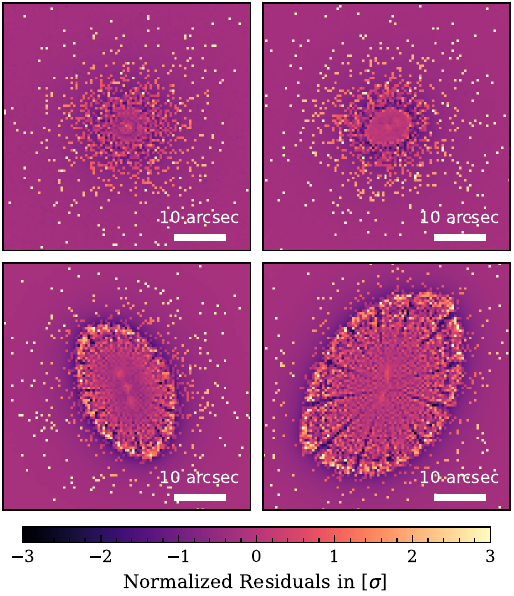} \\
  \includegraphics{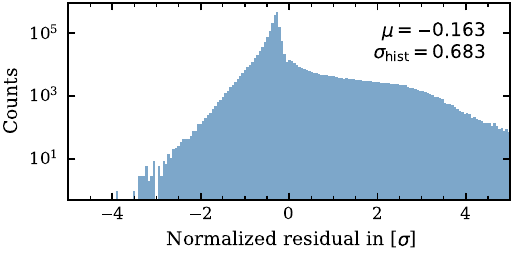}
  \caption{Normalized residuals between \texttt{MARX}-simulated and denoised \acp{PSF} for a subset of \ac{PSF} samples (above). The histogram of normalized residuals for all denoised \acp{PSF} (below).}\label{fig.:psf_diff}
\end{figure}
The denoising of the \ac{PSF} was performed as described in Sec.~\ref{sec.:PSFmodeling}. In order to visualize the quality of the individual fits, we plot the normalized residuals
\begin{equation}
 r_\mathrm{n} = \frac{d-\lambda}{\sqrt{\lambda}}
\end{equation}
in Fig.~\ref{fig.:psf_diff}. These residuals are computed from the \acp{PSF} shown in Fig.~\ref{fig.:psfs_raw} and their corresponding fits shown in Fig.~\ref{fig.:psfs_fit}.
Inspecting Fig.~\ref{fig.:psfs_fit}, we can see that we could recover all the necessary structure from the individual \acp{PSF}. 
In the residual plot we see that for very extended \acp{PSF} we typically cannot fit the outer shells. 
This is due to the used \ac{PSF} model, which favours smooth structures. The shell-like structures, however, have steep gradients, which are disfavoured by the model at use.
The error is small, but clearly structured. 
However, since the amount of flux in these areas is minimal, the error can be neglected. In the Fig.~\ref{fig.:psf_diff} (bottom), we see the distribution of normalized residuals. This shows that they are typically small peaked around zero. 

\subsection{Validation of the spatially variant PSF representation}\label{sec.:psf_valid}
The denoised \acp{PSF} $c_k$ can be used in linear patched convolution (see Sec.~\ref{sec.:fast_psf}) to approximate the true Chandra \ac{PSF} within the \ac{FOV}.
It is legitimate to ask about the error caused by replacing the true \ac{PSF} with an approximated version.
Under the assumption that the true \ac{PSF} $A$ and its approximation by linear patched convolution $A^\star$ are linear operators, we can easily show that the data-space error $\epsilon$ due to the approximation is
\begin{equation}
  \epsilon_i = A_{ij} s_j - A^\star_{ij} s_j = (A_{ij}-A^\star_{ij})s_j \ .
\end{equation}
Setting $s_j = \delta_{jk}$ for a given position $k$ in the unblurred domain extracts the corresponding columns of $A_{ij}$ and $A^\star_{ij}$. 
Since these columns are normalized by construction to fulfill the PSF normalization property, their difference directly yields the relative error at each pixel $i$ in the blurred domain. We denote this error vector as
$\epsilon_i(k)$,
\begin{equation}
  \epsilon_i(k) = (A_{ij}-A^\star_{ij}) \, \delta_{jk} =  (A_{ik}-A^\star_{ik})\ .
\end{equation}
This shows that the error for multiple sources or diffuse structures in the blurred image is a sum of the individual \ac{PSF} approximation errors. To assess the accuracy of the overall reconstruction, we therefore need to compare $A^\star$ and $A$ at all positions. Since we do not have access to the true \ac{PSF} $A$, we will use \texttt{MARX}-simulations as a proxy.

\begin{figure}
  \includegraphics{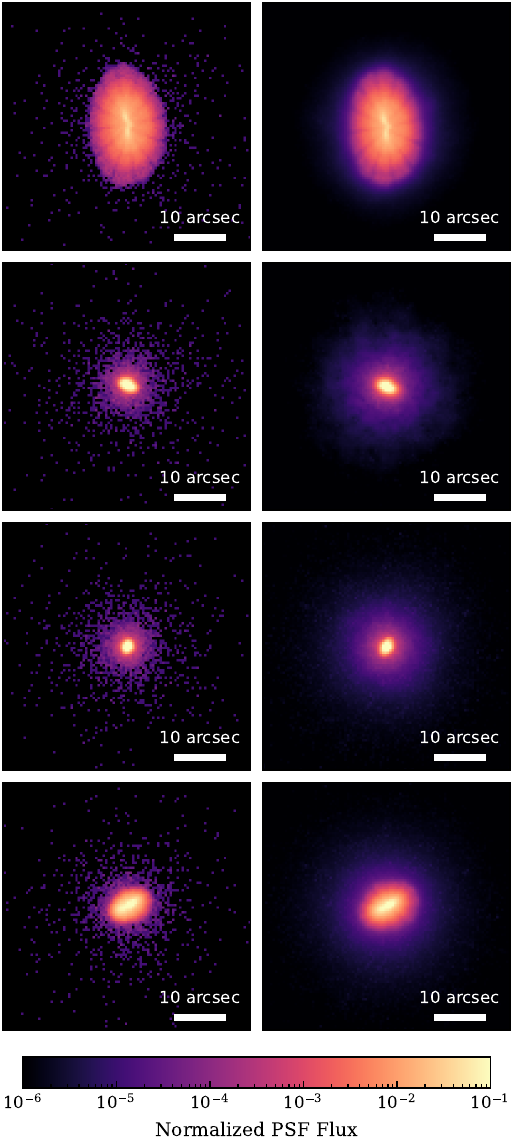}
  \caption{Comparison of additional simulated \acp{PSF} to probe the accuracy of the \ac{PSF} interpolation scheme Sec.~\ref{sec.:fast_psf}. The interpolated \acp{PSF} recover the most important features, size and shape of the simulated \acp{PSF}, as well as some fine details. Residuals are shown in App.~\ref{app.:psfapproximation}.}\label{fig.:psf_approx}
\end{figure}
Testing our $A^\star$ against $A$ for all pixel positions is unfeasible, therefore, we randomly sample positions in this space, simulate the \ac{PSF} with \texttt{MARX}, and compare it to the \ac{PSF} the linear patched convolution approximation yields. 
The normalization of all four contributing \acp{PSF} in linear patched convolution ensures that the resulting \ac{PSF} is always flux-conserving, see Appendix~\ref{app.:flux_cons}.
To compare the approximated \acp{PSF} with the \texttt{MARX}-simulated \acp{PSF} (see Fig.~\ref{fig.:psf_approx}, left panels) we need to correct for the inherent exposure of the latter.
To do so, we normalize the \texttt{MARX}-simulated \acp{PSF} such that they satisfy the \ac{PSF} normalization condition in Eq.~\eqref{eq.:psf_normalization}.
Since the \texttt{MARX}-simulated \acp{PSF} suffer from shot noise, the normalized \acp{PSF} inherit this noise as well.
Poisson distributions are reproductive, which means that the sum $X_s = X_1 + X_2 + ...+ X_n$ of Poisson-distributed random variables $X_1, X_2,..., X_n$ with rates $\lambda_1, \lambda_2, ..., \lambda_n$ is as well Poisson-distributed with rate $\lambda_s = \lambda_1 + \lambda_2+ ...+\lambda_n$. 
Therefore, to estimate the error in the normalization integral, we look at the sum of all pixels, which means that the \ac{S/N} roughly increases with the square root of the number of simulated photons.
This, however, means that the integral flux conservation is already very accurate for the $\mathcal{O}(10^6)$ simulated photons. 
Locally the error can be larger. 

The approximation error $\epsilon$ netween $A$ and $A^\star$ for four very different positions $k$ in the \ac{FOV} can be seen in Fig.~\ref{fig.:psf_approx}.
We clearly see that the overall shape and some fine details of the \acp{PSF} are recovered. 
The difference plots for the \acp{PSF} shown in Fig.~\ref{fig.:psf_approx} can be found in Appendix~\ref{app.:psfapproximation}.

\section{Application to Cassiopeia A}\label{sec.:ChandraApp}
Having demonstrated \ac{NIFTy}’s capability to perform deconvolution and the effectiveness of the linear patched convolution approach in modeling the spatially variant \ac{PSF} of Chandra, we apply the method to a real X-ray observation of Cassiopeia A. This case study serves to validate the framework in a realistic astrophysical context, where complex source morphology and instrumental effects coexist.
\begin{figure*}
\centering
   \includegraphics[width=\textwidth]{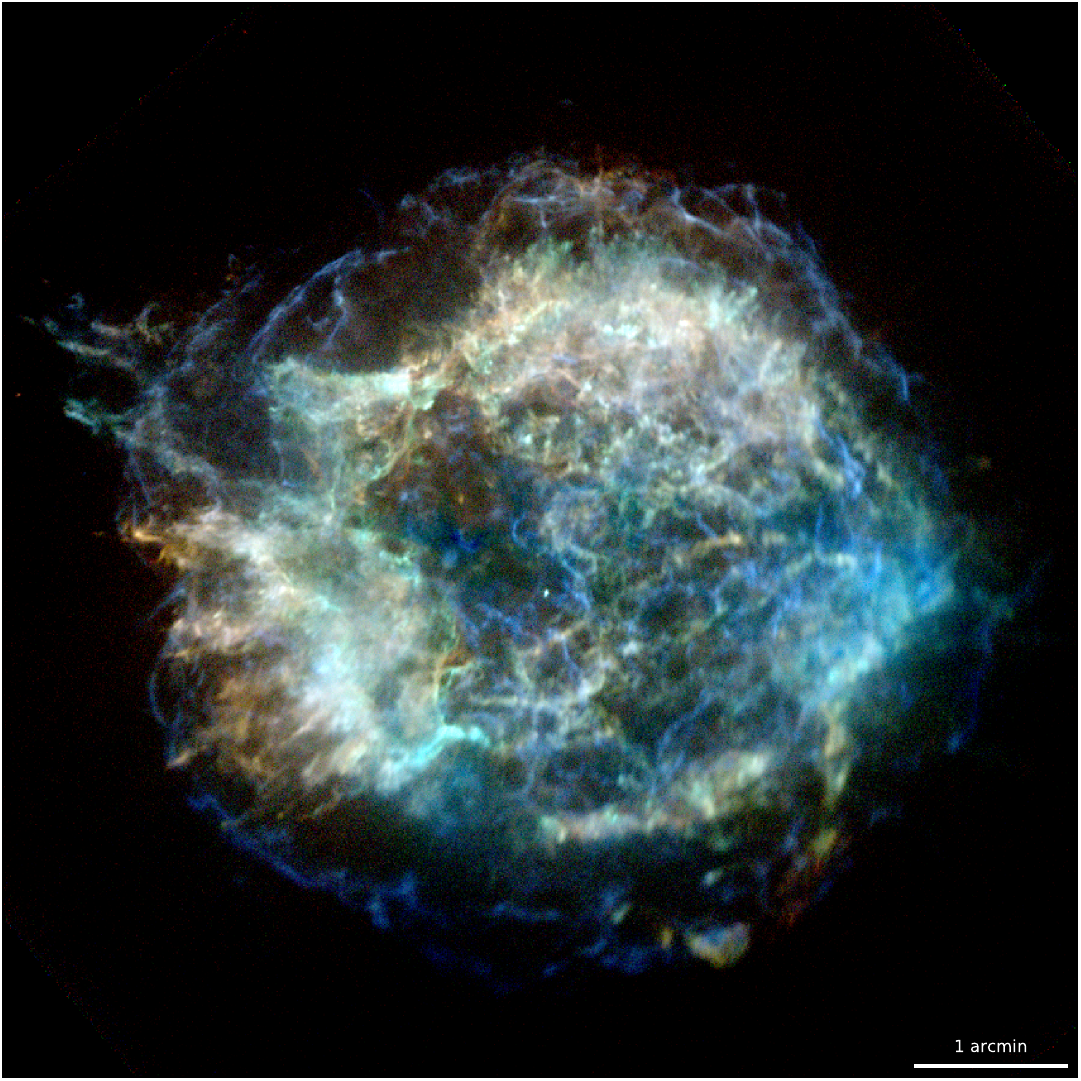}
   \caption{Exposure corrected data~(ObsID:4636) of Cassiopeia A as observed by Chandra. (red: $0.2-1.2 \, \mathrm{keV}$, green: $1.2-2.0 \, \mathrm{keV}$, blue: $2.0-7.0 \, \mathrm{keV}$). One can clearly see the effect of the spatially variant \ac{PSF}. The image is more blurred in the lower right part than in other positions.}\label{fig.:CasA_data}
\end{figure*}
\begin{figure*}
\centering
   \includegraphics[width=\textwidth]{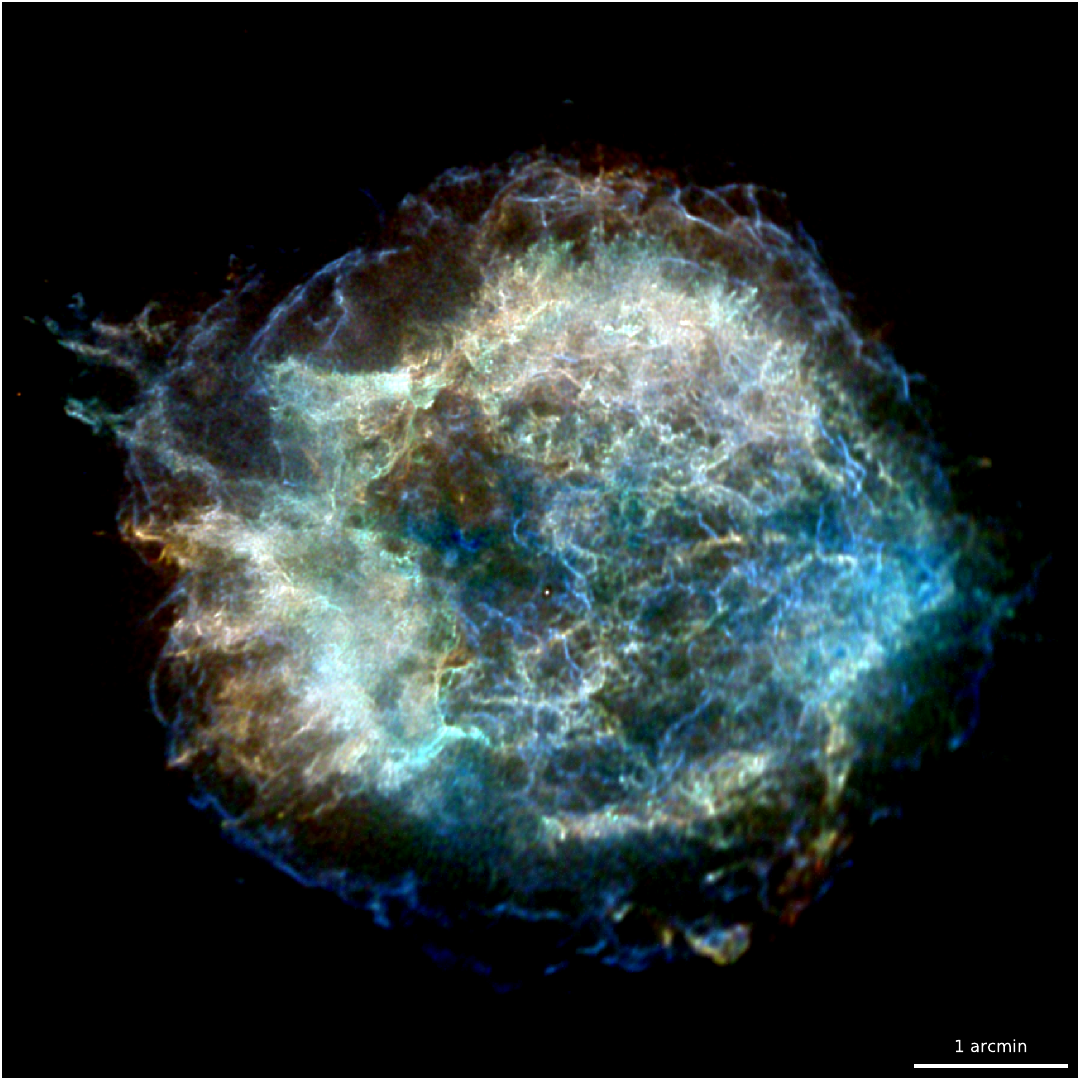}
   \caption{Deblurred signal field $s$ of Cassiopeia A as inferred from~(obsID:4636) taken by Chandra (red:$0.2-1.2 \, \mathrm{keV}$, green: $1.2-2.0 \, \mathrm{keV}$, blue: $2.0-7.0 \, \mathrm{keV}$). The shot noise and blurring are mostly removed. The image gains clarity, especially in the lower right part.}\label{fig.:CasA_inf}
\end{figure*}
\begin{figure*}
\centering
   \includegraphics[width=\textwidth]{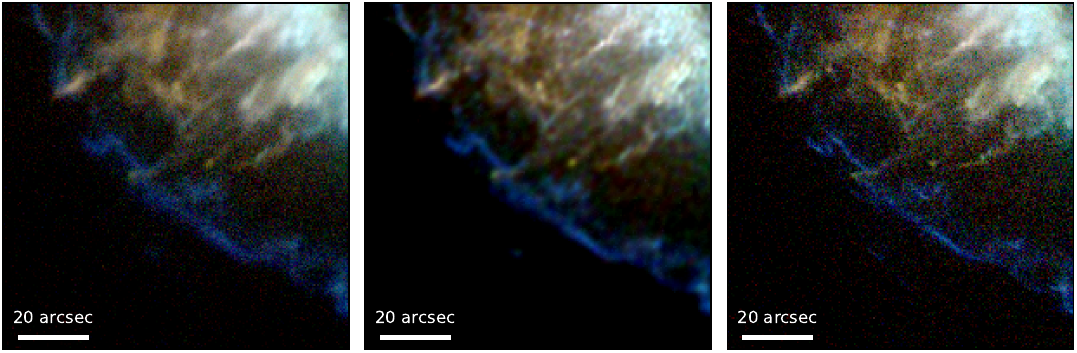}
  \includegraphics[width=\textwidth]{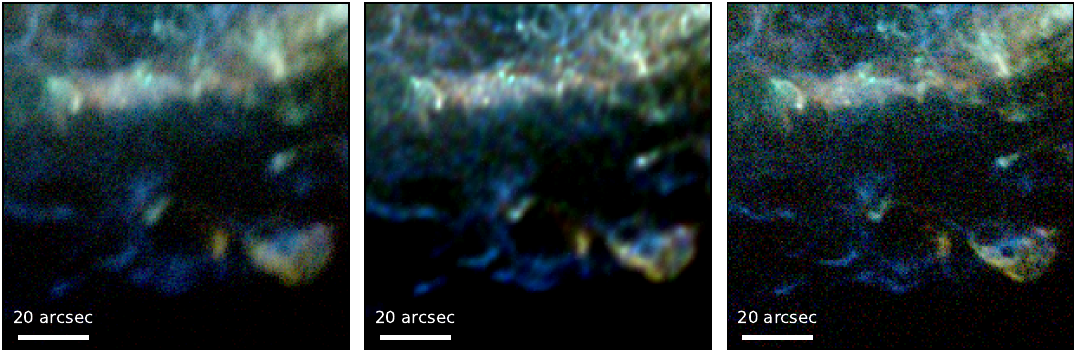}
   \caption{Comparison of observation 4636 (left column), the 4636 reconstruction (middle column), and observation 5196 (right column). The top panels display the lower-left corner of Fig.~\ref{fig.:CasA_inf}. We can clearly see that the reconstructed and deblurred features are real by comparison with the right column, observation 5196. This observation has a different nominal pointing and thus a better resolution in this part of the field of view. The lower panels display the slightly lower-right region of Fig.~\ref{fig.:CasA_inf}. Here too, the resolved structures in the 4636 reconstruction show strong similarity to those in observation 5196.}\label{fig.:CasA-zoom}
\end{figure*}

The data used to demonstrate the method are listed in Table~\ref{tab.:data}. We used \texttt{CIAO}~\citep{fruscione2006ciao,fruscione2026ciao} in version 4.16 to download and read the data, as well as to access the Chandra CALDB.
For the reconstruction, we used only ObsID 4636, while the data from ObsID 5196 served as a reference for validation.
The data were divided into three energy bins: $0.2-1.2 \, \mathrm{keV}$, $1.2-2.0 \, \mathrm{keV}$, and $2.0-7.0 \, \mathrm{keV}$, following the binning of \citet{sakai2023richardson}.
The \ac{FOV} spans approximately $743\arcsec \approx 12.38\arcmin$, with a pixel scale of $\approx 0.5\arcsec$, resulting in $1488 \times 1488$ pixel arrays.
For visualization, we focus on a smaller cut-out of about $850 \times 850$ pixels, corresponding to $\approx 7\arcmin$ on a side. The exposure-corrected data are shown in Fig.~\ref{fig.:CasA_data}.
\subsection{Deblurring Cassiopeia A}
\begin{table}
\centering
\caption{Chandra Observations}\label{tab.:data}
\begin{tabular}{lllll}
\toprule
ObsID & Inst. & R.A. & Decl. & Date \\
\midrule
4636 & ACIS-S & 350.9129 & 58.8412 & 20 Apr 2004\\
5196 & ACIS-I & 350.9129 & 58.7933 & 08 Feb 2004\\
\bottomrule
\end{tabular}
\end{table}

For the reconstruction, we used \ac{NIFTy} to construct a log-normal correlated field prior (see Eq.~\eqref{eq.:log_normal_field}). The specific prior parameters are provided in Appendix~\ref{app.:add_casa}.
The instrument model for Chandra was obtained via \texttt{J-UBIK}~\citep{Eberle2026}, as described in Section~\ref{sec.:Chandra}.
To model the spatially variant \ac{PSF}, we applied the method outlined in Section~\ref{sec.:fast_psf}, using $8\times8$ patches. The \acp{PSF} used in this approach were fitted to \texttt{MARX} simulations (see Sections~\ref{sec.:PSFmodeling} and~\ref{sec.:fit_psf}).

For the inference we used MGVI with 3 antithetic samples. 
The analysis was conducted independently for each energy bin, without imposing additional constraints on the spectral behavior.
The inference converged successfully and yielded a normalized $\chi_\mathrm{norm}^2$ of $1.0$, indicating excellent agreement with the data. 
While incorporating spectral priors would allow access to additional physical information, we deliberately refrained from doing so to isolate the deblurring problem in its purest form.
The results of the three independent inference runs are presented in Fig.~\ref{fig.:CasA_inf}.
Each run took approximately 30 minutes on a single NVIDIA A100 GPU. Additional diagnostic plots are provided in Appendix~\ref{app.:add_casa}.

\subsection{Discussion}
A visual comparison between the observed data (Fig.~\ref{fig.:CasA_data}) and the corresponding inference result (Fig.~\ref{fig.:CasA_inf}) reveals a striking improvement in image quality.
The degradative effects, most notably the spatially variant blurring, have been significantly suppressed in the reconstruction.
In the data, the effect of the spatially variant \ac{PSF} is significant across the field of view: structures in the top-left region appear relatively sharp, while those in the lower-right are strongly blurred. 
This resembles motion blur with a rotation axis near the top-left corner.
The \ac{NIFTy} reconstruction removes this distortion almost entirely, restoring fidelity across the entire field. 
In the reconstruction, we also observe a dark ring surrounding the central point source. 
We identify two reasons for this potential artifact. 
First, for bright point sources pile-up changes the effective shape of the \ac{PSF}. Since we do not include pile-up, the linear \ac{PSF}-model does not capture this behavior, leading to a potential \ac{PSF} misspecification.
Second, to be comparable to \ac{RL}-based reconstruction, we did not include a dedicated point source model in the signal, causing the imperfect modeling of the central point source.
The dark ring is also visible in \citet[Fig.~3]{sakai2023richardson}, where an \ac{RL} algorithm without any prior assumptions is used. Therefore, the structure appears independently of the reconstruction method or any prior assumptions, which suggests that pile-up is the primary cause. 

A detailed inspection via zoom-ins (Fig.~\ref{fig.:CasA-zoom}) further highlights the method's capability.
The first row shows a region in the bottom-left of Cassiopeia A, while the second row focuses on a feature in the lower central area.
In both cases, the reconstruction reveals fine structural details that are absent in the input data.
These recovered features are not spurious. 
They are confirmed by comparison with exposure-corrected data from ObsID 5196, which exhibits a higher intrinsic resolution, due to a different nominal pointing.
The agreement between the \ac{NIFTy} reconstruction and the reference dataset provides strong evidence that the recovered structures are not reconstruction artifacts. 

Additional diagnostic plots, including noise-weighted residuals and normalized uncertainties per energy bin, are provided in Appendix~\ref{app.:add_casa}.
These demonstrate that the model fits the data at a level consistent with Poisson noise.
The residuals exhibit no systematic deviations and appear as uncorrelated, Gaussian-like fluctuations, indicating that the model has captured the essential physical and instrumental features of the observation.
This means that the reconstruction explains the observed data well. The uncertainties display the relative uncertainty of the reconstruction, where bright regions are well constrained, while low-flux regions are relatively unconstrained by the data.

\section{Conclusions}
In this work we demonstrated a new method for the removal of spatially variant \ac{PSF} effects, based on \ac{IFT} and \ac{NIFTy}. Therefore we showed that we can independently solve the two involved problems, removal of blur and representation of spatially variant PSFs, and apply them jointly to an X-ray imaging problem.
The method can be used as a powerful and versatile deblurring algorithm for noisy images, capable of competing with state-of-the-art methods such as \ac{RL} algorithm and its regularized variants.
The flexibility of \ac{NIFTy}’s framework enables the incorporation of complex prior models, including point sources, extended emission structures, and instrumental effects, making it well-suited for realistic astrophysical scenarios.
We further showed that the approximation of the \ac{PSF} using denoised simulated Chandra \acp{PSF} successfully captures the spatial variability of the true Chandra \ac{PSF}, even at large off-axis angles.
This advancement allows for accurate deblurring of Chandra observations under challenging observational conditions, where traditional \ac{PSF} models often fail. 
The combination of robust \ac{PSF} modeling and Bayesian inference within \ac{NIFTy} enables high-fidelity denoising and deblurring of X-ray data, while simultaneously providing full uncertainty quantification, which is a critical advantage in scientific inference.

The additive structure of the joint Hamiltonian, Eq.~\eqref{eq.:hamiltonian_likelihood_more}, further facilitates data fusion across multiple observations. 
This was previously demonstrated for multi-module eROSITA data \cite{eberle2026bayesian}, and the approach can be extended to combine observations from different X-ray observatories, such as Chandra and eROSITA, thereby enhancing the \ac{S/N} and enabling more precise source characterization.
Moreover, the framework opens the door to multi-wavelength data fusion, provided that reliable priors on energy-dependent source properties are available. 
This paves the way for truly multi-instrument, multi-band analyses, where information from different parts of the electromagnetic spectrum is combined to improve reconstruction quality and physical interpretability.
The J-UBIK project \citep{Eberle2026}, an open-source collection of instrumental descriptions and representations for Chandra, eROSITA, and JWST that is fully compatible with \ac{NIFTy}, is already publicly available and supports this vision~\citep{ensslin2025universal}.

The modular design of J-UBIK and \ac{NIFTy} further allows for the inclusion of complex instrumental effects, offering promising directions for future work. 
One such effect is pile-up, a non-linear phenomenon that becomes relevant at high count rates and can bias source morphology and flux estimates. 
Incorporating this effect into the forward model would enhance the physical realism of the reconstruction and enable robust analysis of bright sources in crowded fields or high-flux observations.

While \ac{NIFTy} offers superior accuracy, robustness, and uncertainty quantification, its computational cost is higher than that of classical methods like RL, especially in cases where only deblurring is required and high temporal efficiency is essential. 
For high-\ac{S/N} observations under strict time constraints, \ac{RL} remains a practical and efficient alternative that delivers results within a few seconds. The computation time of \ac{NIFTy}, on the other hand, ranges from a few minutes to several hours, depending on the problem and the device used.
Nevertheless, for low-\ac{S/N} regimes, complex source morphologies, or applications demanding rigorous error propagation, \ac{NIFTy} emerges as a compelling choice.
By extending this methodology with more complex prior models, it would furthermore be possible to perform robust point-source detection and component separation \citep{guardiani2025latent}.

Beyond X-ray astronomy, the methodology presented here is broadly applicable to other fields facing spatially variant deblurring challenges, such as medical imaging, microscopy, and computer vision.
Finally, the potential for blind deconvolution, where the \ac{PSF} is inferred simultaneously from the data, becomes feasible when combining observations from multiple instruments or observing modes. 
This could allow for self-consistent \ac{PSF} reconstruction and deblurring without relying on external calibration, significantly enhancing the autonomy and reliability of future data analysis pipelines.

In summary, the methods presented in this work are a significant step forward in Bayesian image reconstruction for astronomy, offering a flexible, physically motivated, and uncertainty-aware framework that is well-positioned to meet the demands of next-generation X-ray and multi-messenger observatories.
\section{Data availability}
The Chandra observations used in this work are publicly available from the Chandra Data Archive at: 
 \url{https://cda.harvard.edu/chaser/} \\
The code used for all benchmarks and for the Cassiopeia A reconstruction is publicly available at: \url{https://gitlab.mpcdf.mpg.de/ift/spatially-variant-psf}.
\\
The benchmark results and the data products generated for the Cassiopeia A reconstruction are publicly available at: \url{https://doi.org/10.5281/zenodo.22741106}.

\begin{acknowledgements}
  V. Eberle, M. Westerkamp, and M. Guardiani acknowledge support for this research through the project Universal Bayesian Imaging Kit (UBIK, Förderkennzeichen 50OO2103) funded by the Deutsches Zentrum für Luft- und Raumfahrt e.V. (DLR).
  V. Eberle and P. Frank acknowledge funding through the German Federal Ministry of Education and Research for the project ErUM-IFT: Informationsfeldtheorie für Experimente an Großforschungsanlagen (Förderkennzeichen: 05D23EO1).
  V. Eberle acknowledges funding through the German Federal Ministry of Education and Research for the project ErUM-IFT-2: Informationsfeldtheorie für Experimente an Großforschungsanlagen (Förderkennzeichen: 05D25EO1).
  J. Stadler acknowledges funding by the Deutsche Forschungsgemeinschaft (DFG, German Research Foundation) under Germany’s Excellence Strategy – EXC-2094 – 390783311.
  and under its Emmy-Noether Programm (STA 1955/ 1-1, Projektnummer 545534254).
  This research has made use of data obtained from the Chandra Data Archive provided by the Chandra X-ray Center (CXC) and has made use of software provided by the Chandra X-ray Center (CXC) in the application package CIAO.
  The authors used large language models (LLMs), specifically ChatGPT, Claude Sonnet, and Qwen to assist with the language editing and phrasing of parts of this manuscript, as well as with writing of code. All scientific content, results, and conclusions remain the sole responsibility of the authors.
\end{acknowledgements}

\bibliographystyle{aa} 
\bibliography{bibliography}

\begin{thebibliography}{34}
\expandafter\ifx\csname natexlab\endcsname\relax\def\natexlab#1{#1}\fi

\bibitem[{Arras {et~al.}(2019)Arras, Baltac, Ensslin, Frank, Hutschenreuter,
  Knollmueller, Leike, Newrzella, Platz, Reinecke, {et~al.}}]{arras2019nifty5}
Arras, P., Baltac, M., Ensslin, T.~A., {et~al.} 2019, Astrophysics Source Code
  Library, ascl

\bibitem[{Arras {et~al.}(2022)Arras, Frank, Haim, Knollm{\"u}ller, Leike,
  Reinecke, \& En{\ss}lin}]{arras2022variable}
Arras, P., Frank, P., Haim, P., {et~al.} 2022, Nature Astronomy, 6, 259

\bibitem[{Cash(1979)}]{cash1979parameter}
Cash, W. 1979, Astrophysical Journal, Part 1, vol. 228, Mar. 15, 1979, p.
  939-947., 228, 939

\bibitem[{Cooley \& Tukey(1965)}]{cooley1965algorithm}
Cooley, J.~W. \& Tukey, J.~W. 1965, Mathematics of computation, 19, 297

\bibitem[{Davis {et~al.}(2012)Davis, Bautz, Dewey, Heilmann, Houck,
  Huenemoerder, Marshall, Nowak, Schattenburg, Schulz,
  {et~al.}}]{davis2012raytracing}
Davis, J.~E., Bautz, M.~W., Dewey, D., {et~al.} 2012, in Space Telescopes and
  Instrumentation 2012: Ultraviolet to Gamma Ray, Vol. 8443, SPIE, 375--386

\bibitem[{Denis {et~al.}(2015)Denis, Thi{\'e}baut, Soulez, Becker, \&
  Mourya}]{denis2015fast}
Denis, L., Thi{\'e}baut, E., Soulez, F., Becker, J.-M., \& Mourya, R. 2015,
  International Journal of Computer Vision, 115, 253

\bibitem[{Dey {et~al.}(2006)Dey, Blanc-Feraud, Zimmer, Roux, Kam, Olivo-Marin,
  \& Zerubia}]{dey_richardsonlucy_2006}
Dey, N., Blanc-Feraud, L., Zimmer, C., {et~al.} 2006, 69, 260, publisher: Wiley
  Online Library

\bibitem[{Eberle {et~al.}(2022)Eberle, Frank, Stadler, Streit, \&
  Enßlin}]{psf2022005033}
Eberle, V., Frank, P., Stadler, J., Streit, S., \& Enßlin, T. 2022, Physical
  Sciences Forum, 5

\bibitem[{Eberle {et~al.}(2023)Eberle, Frank, Stadler, Streit, \&
  Enßlin}]{e25040652}
Eberle, V., Frank, P., Stadler, J., Streit, S., \& Enßlin, T. 2023, Entropy,
  25

\bibitem[{Eberle {et~al.}(2026{\natexlab{a}})Eberle, Guardiani, Westerkamp,
  Frank, Freyberg, Salvato, \& En{\ss}lin}]{eberle2026bayesian}
Eberle, V., Guardiani, M., Westerkamp, M., {et~al.} 2026{\natexlab{a}},
  Astronomy \& Astrophysics, 706, A229

\bibitem[{Eberle {et~al.}(2024)Eberle, Guardiani, Westerkamp, Frank,
  R{\"u}stig, Stadler, \& En{\ss}lin}]{eberle2024j}
Eberle, V., Guardiani, M., Westerkamp, M., {et~al.} 2024, arXiv preprint
  arXiv:2409.10381

\bibitem[{Eberle {et~al.}(2026{\natexlab{b}})Eberle, Guardiani, Westerkamp,
  Frank, Rüstig, Stadler, \& Enßlin}]{Eberle2026}
Eberle, V., Guardiani, M., Westerkamp, M., {et~al.} 2026{\natexlab{b}}, Journal
  of Open Source Software, 11, 7768

\bibitem[{Edenhofer {et~al.}(2024)Edenhofer, Frank, Roth, Leike, Guerdi,
  Scheel-Platz, Guardiani, Eberle, Westerkamp, \& Enßlin}]{niftyre}
Edenhofer, G., Frank, P., Roth, J., {et~al.} 2024, Journal of Open Source
  Software, 9, 6593

\bibitem[{En{\ss}lin {et~al.}(2025)En{\ss}lin, Eberle, Guardiani, \&
  Westerkamp}]{ensslin2025universal}
En{\ss}lin, T., Eberle, V., Guardiani, M., \& Westerkamp, M. 2025, arXiv
  preprint arXiv:2512.12376

\bibitem[{En\ss{}lin {et~al.}(2009)En\ss{}lin, Frommert, \&
  Kitaura}]{ensslin2009}
En\ss{}lin, T.~A., Frommert, M., \& Kitaura, F.~S. 2009, Phys. Rev. D, 80,
  105005

\bibitem[{Fruscione {et~al.}(2006)Fruscione, McDowell, Allen, Brickhouse,
  Burke, Davis, Durham, Elvis, Galle, Harris, {et~al.}}]{fruscione2006ciao}
Fruscione, A., McDowell, J.~C., Allen, G.~E., {et~al.} 2006, in Observatory
  Operations: Strategies, Processes, and Systems, Vol. 6270, SPIE, 586--597

\bibitem[{Fruscione {et~al.}(2026)Fruscione, McDowell, Burke,
  Cresitello-Dittmar, Evans, Evans, Glotfelty, G{\"u}nther, Huenemoerder, Joye,
  {et~al.}}]{fruscione2026ciao}
Fruscione, A., McDowell, J.~C., Burke, D.~J., {et~al.} 2026, The Astrophysical
  Journal, 1005, 116

\bibitem[{Guardiani {et~al.}(2025)Guardiani, Eberle, Westerkamp, R{\"u}stig,
  Frank, \& En{\ss}lin}]{guardiani2025latent}
Guardiani, M., Eberle, V., Westerkamp, M., {et~al.} 2025, Astronomy \&
  Astrophysics, 703, A203

\bibitem[{Knollm{\"u}ller \& En{\ss}lin(2019)}]{knollmuller2019metric}
Knollm{\"u}ller, J. \& En{\ss}lin, T.~A. 2019, arXiv preprint arXiv:1901.11033

\bibitem[{Lauer(2002)}]{lauer2002deconvolution}
Lauer, T. 2002, in Astronomical data analysis II, Vol. 4847, SPIE, 167--173

\bibitem[{Lucy(1974)}]{lucy1974iterative}
Lucy, L.~B. 1974, Astronomical Journal, Vol. 79, p. 745 (1974), 79, 745

\bibitem[{Nagy \& O'Leary(1997)}]{nagy1997fast}
Nagy, J.~G. \& O'Leary, D.~P. 1997, in Advanced Signal Processing: Algorithms,
  Architectures, and Implementations VII, Vol. 3162, SPIE, 388--399

\bibitem[{Pumpe {et~al.}(2018)Pumpe, Reinecke, \&
  En{\ss}lin}]{pumpe2018denoising}
Pumpe, D., Reinecke, M., \& En{\ss}lin, T.~A. 2018, Astronomy \& Astrophysics,
  619, A119

\bibitem[{Ramos(2024)}]{ramos2024solar}
Ramos, A.~A. 2024, Astronomy \& Astrophysics, 688, A88

\bibitem[{Richardson(1972)}]{richardson1972bayesian}
Richardson, W.~H. 1972, JoSA, 62, 55

\bibitem[{Rudin {et~al.}(1992)Rudin, Osher, \& Fatemi}]{rudin1992nonlinear}
Rudin, L.~I., Osher, S., \& Fatemi, E. 1992, Physica D: nonlinear phenomena,
  60, 259

\bibitem[{Sakai {et~al.}(2023)Sakai, Yamada, Sato, Hayakawa, Higurashi, \&
  Kominato}]{sakai2023richardson}
Sakai, Y., Yamada, S., Sato, T., {et~al.} 2023, The Astrophysical Journal, 951,
  59

\bibitem[{Scheel-Platz {et~al.}(2023)Scheel-Platz, Knollm{\"u}ller, Arras,
  Frank, Reinecke, J{\"u}stel, \& En{\ss}lin}]{scheel2023multicomponent}
Scheel-Platz, L., Knollm{\"u}ller, J., Arras, P., {et~al.} 2023, Astronomy \&
  Astrophysics, 680, A2

\bibitem[{Selig {et~al.}(2013)Selig, Bell, Junklewitz, Oppermann, Reinecke,
  Greiner, Pachajoa, \& En{\ss}lin}]{selig2013nifty}
Selig, M., Bell, M.~R., Junklewitz, H., {et~al.} 2013, Astronomy \&
  Astrophysics, 554, A26

\bibitem[{Selig \& En{\ss}lin(2015)}]{selig2015denoising}
Selig, M. \& En{\ss}lin, T.~A. 2015, Astronomy \& Astrophysics, 574, A74

\bibitem[{Selig {et~al.}(2015)Selig, Vacca, Oppermann, \&
  En{\ss}lin}]{selig2015denoised}
Selig, M., Vacca, V., Oppermann, N., \& En{\ss}lin, T.~A. 2015, Astronomy \&
  Astrophysics, 581, A126

\bibitem[{Thi{\'e}baut {et~al.}(2016)Thi{\'e}baut, D{\'e}nis, Soulez, \&
  Mourya}]{thiebaut2016spatially}
Thi{\'e}baut, {\'E}., D{\'e}nis, L., Soulez, F., \& Mourya, R. 2016, in
  Adaptive Optics Systems V, Vol. 9909, International Society for Optics and
  Photonics, 99097N

\bibitem[{Wang {et~al.}(2004)Wang, Bovik, Sheikh, \&
  Simoncelli}]{wang2004image}
Wang, Z., Bovik, A.~C., Sheikh, H.~R., \& Simoncelli, E.~P. 2004, IEEE
  transactions on image processing, 13, 600

\bibitem[{Westerkamp {et~al.}(2024)Westerkamp, Eberle, Guardiani, Frank, Platz,
  Arras, Knollm{\"u}ller, Stadler, \& En{\ss}lin}]{westerkamp2024sn1006}
Westerkamp, M., Eberle, V., Guardiani, M., {et~al.} 2024, Astronomy \&
  Astrophysics, 684, A155

\end{thebibliography}

\begin{appendix}
  \acresetall
\section{Additional details on CasA inference}\label{app.:add_casa}

Here, we want to show additional information and metrics of the inference of Cassiopeia A. Tab.~\ref{tab:nifty_para} shows the hyperparameters for the correlated field model used. Fig.~\ref{fig.:post_mean_casa} shows the posterior mean for each energy bin. This shows the three energy bins of Fig.~\ref{fig.:CasA_inf} individually, including a colorbar. The pixel wise standard deviation between the posterior samples for each energy bin is shown in Fig.~\ref{fig.:post_std_casa}. The noise-weighted residuals,
\begin{equation}
  r_\mathrm{nwr} = \frac{1}{N_\mathrm{s}} \sum_i^{N_\mathrm{s}} \frac{d - \mathcal{R}(s_i)}{\sqrt{\mathcal{R}(s_i)}}
\end{equation}
are shown in Fig.~\ref{fig.:nwr_casa}. The noise-weighted residuals show that the remaining structure in the exposed and blurred data space is mostly noise. One can spot the bright point source in the center. Since the brightness of this point source is not correlated with the surrounding structure, it is necessary to include an uncorrelated component in the forward model. As this is not possible in a standard RL method, we deliberately did not include this extra component, although it could be straightforwardly incorporated within the NIFTy framework.
\begin{figure}[h!]
  \includegraphics{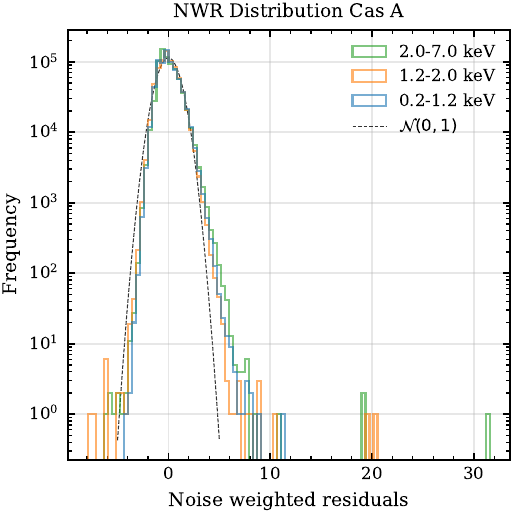}
  \caption{Histogram of the noise-weighted residuals for different energy bins. The residuals follow approximately a normal distribution. 
  The plot also shows the expected right skew in low-flux regions.}
\end{figure}

\begin{figure}[h!]
  \includegraphics[width=0.49\textwidth]{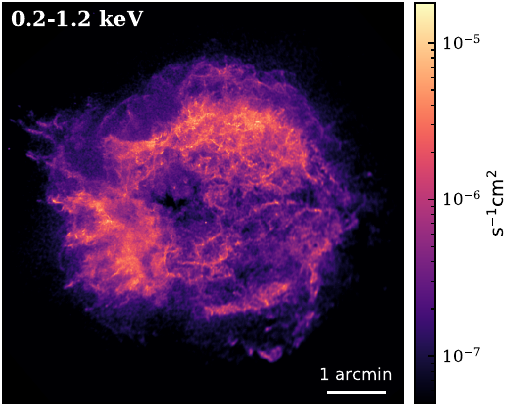}\\
  \includegraphics[width=0.49\textwidth]{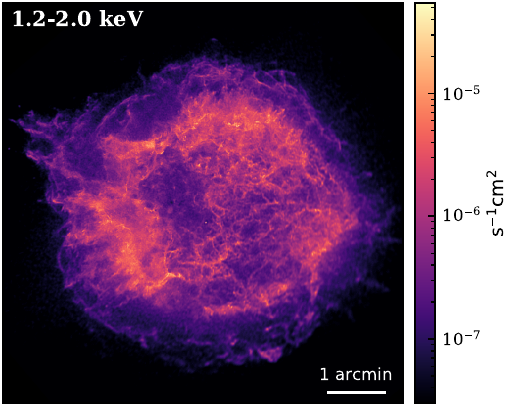}\\
  \includegraphics[width=0.49\textwidth]{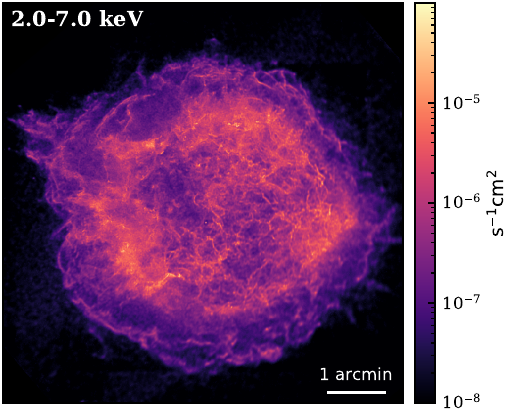}
  \caption{Posterior mean sky brightness in [$\mathrm{s^{-1} cm^{-2}}$] }\label{fig.:post_mean_casa}
\end{figure}

\begin{figure}[h!]
  \includegraphics[width=0.49\textwidth]{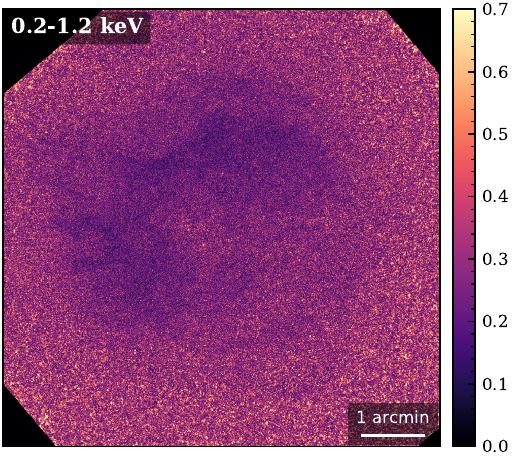}\\
  \includegraphics[width=0.49\textwidth]{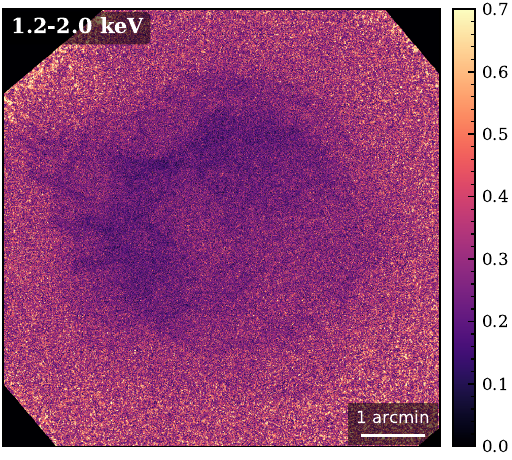}\\
  \includegraphics[width=0.49\textwidth]{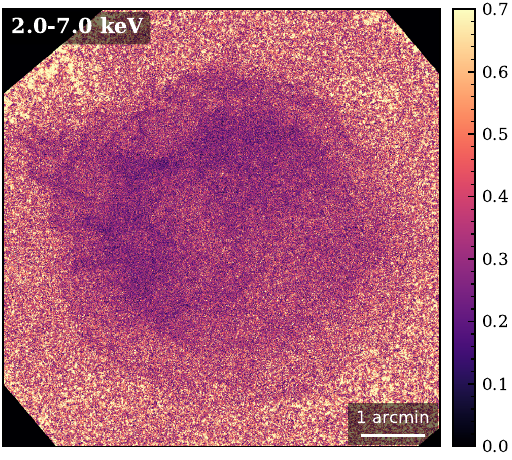}
  \caption{Posterior standard deviation divided by the posterior mean sky brightness, Fig.~\ref{fig.:post_mean_casa}.}\label{fig.:post_std_casa}
\end{figure}

\begin{figure}[h!]
  \includegraphics[width=0.49\textwidth]{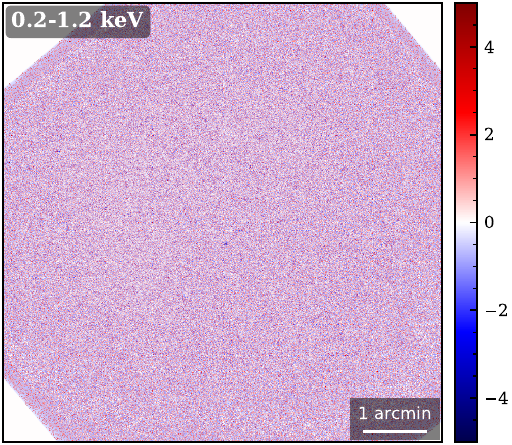}\\
  \includegraphics[width=0.49\textwidth]{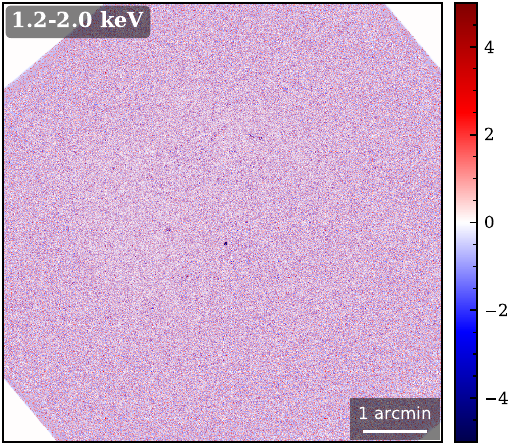}\\
  \includegraphics[width=0.49\textwidth]{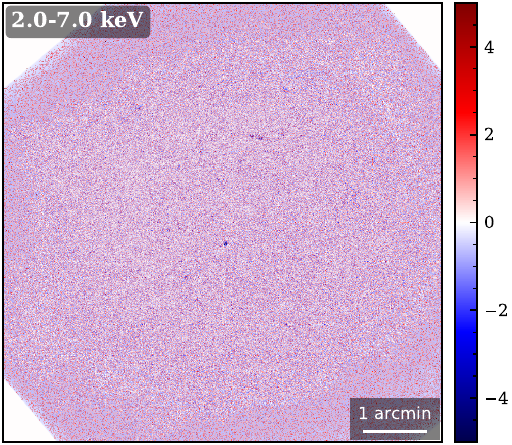}
  \caption{Noise-weighted residuals of, Fig.~\ref{fig.:post_mean_casa}}\label{fig.:nwr_casa}
\end{figure}

\begin{table}[h!]
\small
\caption{Hyperparameters used for the correlated field model.}
\label{tab:nifty_para}
\centering
\begin{tabular}{llp{3.9cm}}
\toprule
Parameter               & ($\mu$ , $\sigma$ )     &  Description\\
\midrule
offset mean              & - 16      &  zeromode\\
offset std               & (1, 1)    &  (Std, Std of the Std) of the zeromode\\
fluctuations               & (1, 0.5)    & describes the amplitude of $\phi$ \\
loglog average slope      & (-2.3, 0.5)    & slope of the power law of the power spectrum of $\phi$ \\
flexibility             & None    &  \\
asperity             & None    &  \\
\bottomrule
\end{tabular}
\end{table}

\clearpage
\section{MARX}\label{app.:marx}
We used \texttt{MARX} to simulate all the \acp{PSF} required for the spatially variant deblurring approach described in this work.
Therefore, we divide the image into $n_\mathrm{patch}$ patches. 
Given some observation with a specific nominal pointing and a image center in sky coordinates, which we define beforehand, we calculate the sky coordinate for each patch center and simulate the observation of a point source, the \ac{PSF}, at this position. 
For the simulation, we assume a flat source spectrum for each energy bin. 
\begin{table}[h!]
\small
\caption{Simulation parameters used for the \texttt{MARX} ray-tracing simulations.}
\label{tab:marx_params}
\centering
\begin{tabular}{llp{3.9cm}}
\toprule
Parameter & Value & Description \\
\midrule
\texttt{SourceType}               & \texttt{POINT}       & Type of simulated source \\
\texttt{SpectrumType}             & \texttt{FLAT}        & Spectral model of source \\
\texttt{SourceFlux}               & $10^{-3}$            & Source flux (photons\,s$^{-1}$\,cm$^{-2}$) \\
\texttt{ExposureTime}             & 0.0                  & Exposure time (s) \\
\texttt{NumRays}                  & $10^{6}$             & Number of simulated rays \\
\texttt{RA\_Nom}                  & from data            & Nominal pointing R.A. (deg) \\
\texttt{Dec\_Nom}                 & from data            & Nominal pointing Dec. (deg) \\
\texttt{Roll\_Nom}                & from data            & Nominal roll angle (deg) \\
\texttt{DetOffsetX}               & from data            & Detector offset in $X$ (mm) \\
\texttt{DetOffsetZ}               & from data            & Detector offset in $Z$ (mm) \\
\texttt{AspectBlur}               & 0.25                 & Aspect blur (arcsec) \\
\texttt{DetectorType}             & \texttt{ACIS-S}      & Detector type \\
\texttt{DetIdeal}                 & \texttt{yes}         & Use ideal detector response \\
\texttt{GratingType}              & \texttt{NONE}        & Grating type \\
\texttt{MirrorType}               & \texttt{HRMA}        & Mirror type \\
\texttt{HRMA\_Ideal}              & \texttt{no}          & Use ideal HRMA model \\
\texttt{HRMAVig}                  & 1.0                  & HRMA vignetting factor \\
\texttt{DitherModel}              & \texttt{INTERNAL}    & Dither model \\
\texttt{Verbose}                  & \texttt{no}          & Verbose output \\
\texttt{ACIS\_FTT}\tablefootmark{a}  & 0.000             & ACIS frame transfer time (s) \\
\texttt{HRMA\_Use\_Struts}        & \texttt{yes}         & Include HRMA struts \\
\texttt{DetExtendFlag}            & \texttt{yes}         & Extend detector \\
\texttt{MinEnergy}                & $E_\mathrm{min}$     & Minimum energy of bin (keV) \\
\texttt{MaxEnergy}                & $E_\mathrm{max}$     & Maximum energy of bin (keV) \\
\bottomrule
\end{tabular}
\tablefoottext{a}{\texttt{ACIS\_FTT} = \texttt{ACIS\_Frame\_Transfer\_Time}}
\end{table}

\section{Derivation of Richardson-Lucy}\label{app.:regularization}
In the section we will use the following mathematical symbols:
\begin{itemize}
  \item $\cdot$ is a point-wise product, also called Hadamard product.
  \item $\ast$ is a convolution, i.e., $\sum_l a_{k-l}b_l = (a\ast b)_k $. 
  \item $\lVert a \rVert_2 = \sqrt{\sum_ia_i^2}$ is the 2-norm, or Euclidian norm. 
  \item $\lVert\nabla a\rVert_{2,1} = \sum_i \sqrt{(\nabla_x a_i)^2+(\nabla_y a_i)^2}$ is a TV-norm, or (2,1)-norm.
  The (2, 1)-norm is calculated by first taking the 2-norm of the gradient vector at each pixel. After that we evaluate the 1-norm of the resulting magnitudes over all pixels.
\end{itemize}
\subsection{Richardson-Lucy}
First we want to show how to get the \ac{RL} algorithm from the Poisson likelihood. Therefore we use the likelihood Hamiltonian, take the derivative with respect to s and set it to zero.
\begin{align}
\mathcal{H}(d \mid s) &= \sum_i \left( \lambda_i - d_i \log \lambda_i + \log d_i! \right) \\ 
  \mathcal{H}(d \mid s) &= \sum_i  (a\ast s)_i - d_i \log (a\ast s)_i + \log d_i! \\
    \frac{\partial \mathcal{H}(d \mid s)}{\partial s} &= \sum_i \frac{\partial(a \ast s)_i}{\partial s_j} \cdot \left(1 - \frac{d_i }{(a\ast s)_i}\right) \label{eq.:PureRL_derivative}
\end{align}
Explicitly writing down the convolution:
\begin{equation}
  \frac{\partial(a \ast s)_i}{\partial s_j} 
  = \frac{\partial\sum_k a_{i-k}s_k}{\partial s_j} = a_{i-j} .
\end{equation}
Putting this into Eq.~\eqref{eq.:PureRL_derivative} gives:
\begin{align}
    \frac{\partial \mathcal{H}(d \mid s)}{\partial s_j}
    &=  \sum_i a_{i-j} \cdot \left(1 - \frac{d_i }{(a\ast s)_i}\right) \\
    &= \sum_i \tilde{a}_{j-i} \cdot \left(1 - \frac{d_i }{(a\ast s)_i}\right) \\
    &=  \tilde{a} \ast \left(1 - \frac{d}{(a\ast s)}\right) \ . \label{eq.:PureRL_derivative_2}
\end{align}
Since we are now summing over the first index of $a_{i-j}$ and not over the second index, this is not a convolution with $a$, but a convolution with the mirrored \ac{PSF}, $\tilde{a}_{j-i} = a_{i-j}$. Equivalently, since $a$ is real-valued, one can take the complex conjugate of the Fourier transformed \ac{PSF} $a$ to get $\tilde{a}$.

If we set Eq.~\eqref{eq.:PureRL_derivative_2} to zero and separate the two terms in the parentheses and use that $\tilde{a}\ast 1 = 1$ for normalized \acp{PSF}, we get.
\begin{equation}
      1=  \tilde{a} \ast \frac{d}{(a\ast s)}
\end{equation}
In the next step we point-wise multiply both sides with $s$ and then use fixed-point iteration, and state that
\begin{equation}
  s^{(n+1)} = s^{(n)}\cdot \left(\frac{d}{(a\ast s^{(n)})}\ast \tilde{a}\right) \eqqcolon s^{(n)} \cdot c^{(n)}\ .
\end{equation}

\subsection{TV regularization}
As in the \ac{RL} algorithm, we want to minimize the Hamiltonian, now including the \ac{TV} regularization term weighted by the parameter $\mu$,
\begin{align}
\mathcal{H}_\mathrm{TV}(d,s) = \sum_i \left( \lambda_i - d_i \log \lambda_i + \log d_i! \right) \  + \mu \, \lVert \nabla s \rVert_1 \ .
\end{align}
Again, we take the derivative with respect to $s$, which yields the same term as in the plain RL case, plus an additional term due to the \ac{TV} regularization,
\begin{align}\label{eq.:RL_TV_step}
  \frac{\partial \mathcal{H}_\mathrm{TV}}{\partial s} &= \tilde{a} \ast \left(1 - \frac{d_i }{(a\ast s)_i}\right) + \mu\frac{\partial\ \lVert \nabla s \rVert_1}{\partial s} \,.
\end{align}
The derivative of the regularization term with respect to $s$ turns into
\begin{equation}
  \frac{\partial \lVert \nabla s \rVert_1}{\partial s} = -\mathrm{div}\left(\frac{\nabla s}{\lvert \nabla s \rvert}\right) \ .
\end{equation}
Now we use that result in the Eq.~\eqref{eq.:RL_TV_step} and set it to zero,
\begin{align}
  \frac{\partial \mathcal{H}_\mathrm{TV}}{\partial s} &= \tilde{a} \ast \left(1 - \frac{d_i }{(a\ast s)_i}\right) - \mu \, \mathrm{div}\left(\frac{\nabla s}{\lvert \nabla s \rvert}\right) = 0
\end{align}
In the next step we separate the terms and use that $\tilde{a} \ast 1 = 1$, due to \ac{PSF} normalization,
\begin{align}
  1  &=  \frac{\tilde{a} \ast \frac{d}{a\ast s}}{\tilde{a}\ast 1 - \mu \, \mathrm{div}\left(\frac{\nabla s}{\lvert \nabla s \rvert}\right)} \\
    &= \frac{\tilde{a} \ast \frac{d}{a\ast s}}{1 - \mu \, \mathrm{div}\left(\frac{\nabla s}{\lvert \nabla s \rvert}\right)}\ .
\end{align}
Then, as in RL algorithm we set this up as a fixed-point iteration, by multiplying with $s$, and obtain
  \begin{align}
s^{(n+1)} &= s^{(n)} \cdot \left(\frac{d}{\left(a\ast s^{(n)}\right)}\ast \tilde{a}\right) \cdot \left[1 - \mu \, \mathrm{div} \, \left(\frac{\nabla s^{(n)}}{\lvert\nabla s^{(n)}\rvert}\right)\right]^{-1} \\ 
& \eqqcolon s^{(n)} \cdot \frac{c^{(n)}}{r^{(n)}} \ .
\end{align}
In practice, we clip the denominator $r^{(n)}$ to a small positive minimum and regularize $|\nabla s|$ as $\sqrt{|\nabla s|^2 + \epsilon}$ to avoid numerical instabilities. Both the clipping value and $\epsilon$ are set to $10^{-14}$.

\subsection{Tikhonov regularization}
Here, we want to show how to derive the iterative formula from the joint Hamiltonian, consisting of the Poisson likelihood and the Tikhonov regularization term,
\begin{align}
\mathcal{H}_\mathrm{TIK}(d,s) = \sum_i \left( \lambda_i - d_i \log \lambda_i + \log d_i! \right) \  + \mu \, \lVert \nabla s \rVert_2^2 \ .
\end{align}
As in the derivation of RL without regularization, we start with the Hamiltonian, take the derivative with respect to $s$, and set it to zero. The Poisson likelihood term stays the same, but we get an additional term for the regularization,
\begin{align}
  \frac{\partial \mathcal{H}_\mathrm{TIK}}{\partial s} &= \tilde{a} \ast \left(1 - \frac{d_i }{(a\ast s)_i}\right) + \mu\ \frac{\partial\lVert \nabla s \rVert_2^2}{\partial s} \, .
\end{align}
It can be shown that 
\begin{equation}
  \frac{\partial \lVert \nabla s \rVert_2^2}{\partial s} = -2 \Delta s \,,
\end{equation}
which leaves us with,
\begin{equation}
  \frac{\partial \mathcal{H}_\mathrm{TIK}}{\partial s} = \tilde{a} \ast \left(1 - \frac{d_i }{(a\ast s)_i}\right)  + 2\mu (- \Delta s) \ .
\end{equation}
We again set this to zero.
Under the assumption of a normalized \ac{PSF}, we have  $\tilde{a} \ast 1 = 1$. 
Since $\mu$ is an arbitrarily chosen parameter, we can absorb the factor of 2 into $\mu$, denoted as $\mu_\star$,
\begin{align}
  1  &=  \frac{\tilde{a} \ast \frac{d}{a\ast s}}{\tilde{a}\ast 1 + 2 \mu (-\Delta s)} \\
    &= \frac{\tilde{a} \ast \frac{d}{a\ast s}}{1+ \mu_{\star} (-\Delta s)}\ .
\end{align}
To avoid numerical instabilities from evaluating $-\Delta s^{(n)}$ point-wise, we instead solve 
\begin{equation}
  (1-\mu_\star\Delta)s^{(n+1)} = s^{(n)}\cdot c^{(n)}
\end{equation}
efficiently in Fourier space, where the Laplacian becomes a multiplication by $-|k|^2$,
\begin{equation}
 s^{(n+1)} = \mathcal{F}^{-1}\left[ \left(1 + \mu_\star \lvert k\rvert^2\right)^{-1}\mathcal{F}\left(s^{(n)} \cdot c^{(n)}\right)\right] \, .
\end{equation}
For brevity, we omit the subscript $\star$ in the main text.
This filter, acting as a Wiener-like damping of high spatial frequencies, avoids the numerical instabilities of the naive explicit update.

\section{Flux conservation in patched convolution}\label{app.:flux_cons}

The order of operations defines whether the flux is truly conserved. In scenario A, we weight the signal $s$ and then convolve it:
\begin{equation}
  \int s^\prime(y) \, \mathrm{d}y = \sum_k \int \mathrm{d}y \int \mathrm{d}x \, c_k(y-x)w_k(x)s(x) \, .
\end{equation}
Here, we can change the order of integration and get
\begin{equation}
  \int s^\prime(y) \, \mathrm{d}y = \int \mathrm{d}x \sum_k \int \mathrm{d}y \, c_k(y-x)w_k(x)s(x) \, .
\end{equation}
Due to the normalization of the \ac{PSF}, $\int \mathrm{d}y \, c_k(y-x) = 1$, we get
\begin{equation}
  \int s^\prime(y) \, \mathrm{d}y = \int \mathrm{d}x \sum_k w_k(x)s(x)
\end{equation}
and due t to the definition of the weights, $\sum_k w_k(x) = 1$, we are left with
\begin{equation}
  \int s^\prime(y) \, \mathrm{d}y = \int \mathrm{d}x \ s(x) \ ,
\end{equation}
meaning that the flux is conserved.

In scenario B, we first convolve and then weight, giving
\begin{equation}
  \int s^\prime(y) \, \mathrm{d}y = \sum_k \int \mathrm{d}y \ w_k(y)\int \mathrm{d}x \, c_k(y-x)s(x) \ .
\end{equation}
After swapping the order of integration we obtain
\begin{equation}
  \int s^\prime(y) \, \mathrm{d}y = \int \mathrm{d}x \, s(x) \sum_k\int \mathrm{d}y \, c_k(y-x) w_k(y) \ .
\end{equation}
This means that in order to conserve the flux, the sum of convolutions of the weights and the kernels must equal one,
\begin{equation}
  \sum_k(w_k \ast c_k)(x)=1 \quad \forall x \ ,
\end{equation}
which is harder to satisfy compared to scenario A, as it imposes additional constraints on the $w_k$'s.

The physical difference is that in scenario A we weight the true flux, whereas in scenario B we weight the blurred flux. Given the constraint $\sum_k w_k(x) = 1$, an intuitive interpretation is that the true flux is distributed into multiple channels, each of which is independently blurred and subsequently summed. This ensures that the total incoming flux is conserved and only redistributed by the blurring.

\section{Further deconvolution metrics}\label{app.:deconv_metrics}
\begin{figure}
  \includegraphics[width=0.49\textwidth]{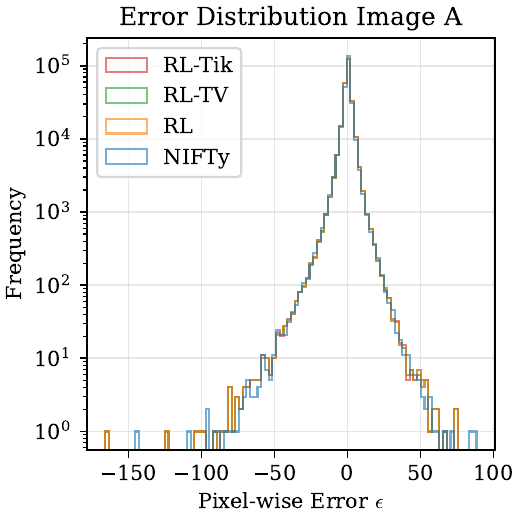}\\
  \includegraphics[width=0.49\textwidth]{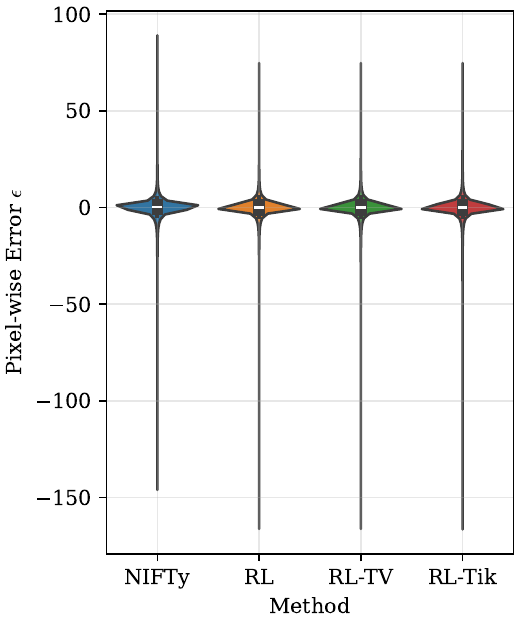} 
  \caption{Error $\epsilon$ (see Eq.~\ref{eq.:error_method}) distributions of four approximation methods reconstructing image A. Pixel-wise residuals are shown as histograms (top panel). The violin plots (lower panel) depict the distribution, with the box plots within the violins indicating the interquartile range (IQR), from 25th to 75th percentile. The line within the box represents the median.}\label{fig.:ImageAHist}
\end{figure}

\begin{figure}
  \includegraphics[width=0.49\textwidth]{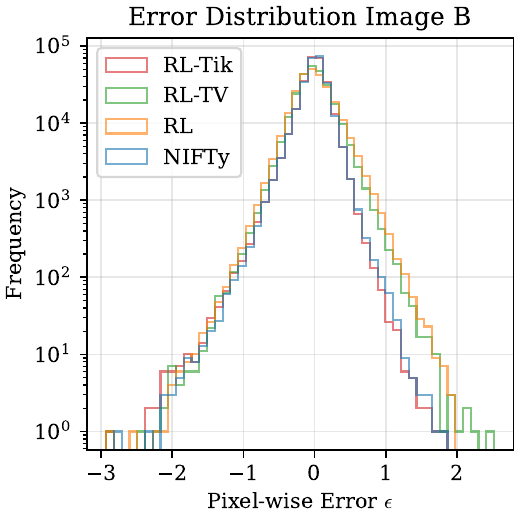}\\
  \includegraphics[width=0.49\textwidth]{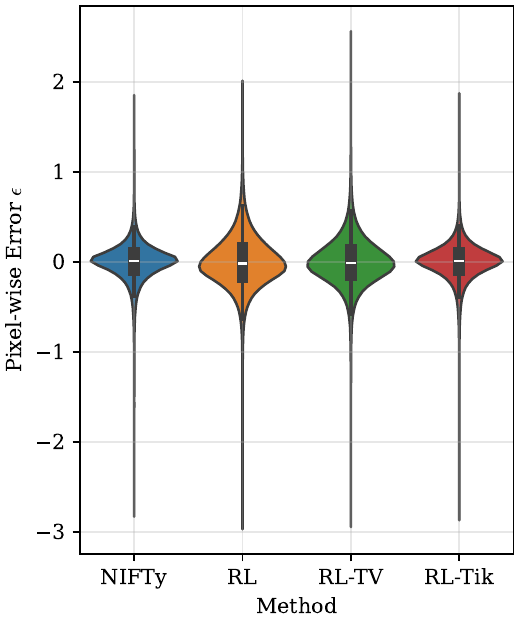} 
  \caption{As Fig.~\ref{fig.:ImageAHist} but for reconstruction image B.}\label{fig.:ImageBHist}
\end{figure}

Here, we show additional metrics for the deconvolution comparison of the four different algorithms, plain \ac{RL}, \ac{RL} with \ac{TV} regularization, \ac{RL} with Tikhonov regularization and \ac{NIFTy}.
In Figs.~\ref{fig.:ImageAHist} and~\ref{fig.:ImageBHist}, we see the error distributions for the different methods. Additionally to the histogram, we get information about the median, the interquartile range (IQR) and the density estimation of the distribution from the violin plots in the same figure. Here, the errors are defined as
\begin{equation}\label{eq.:error_method}
  \epsilon = s_{\mathrm{method}}-s_{\mathrm{gt}} \ ,
\end{equation}
where $s_{\mathrm{gt}}$ denotes the ground truth of the signal and $s_{\mathrm{method}}$ the signal estimate of a method.

For image A, we can see that the errors look quite similarly distributed, however, the distribution is a bit less spread out for \ac{NIFTy}. In the violin plot we see that NIFTy seems to be slightly biased towards positive errors, while the other methods are more negatively biased.
For image B, we see the same behavior, but clearer. This is due to the lower-\ac{S/N} (see Fig.~\ref{fig.:snrmaps}). Here, plain \ac{RL} and \ac{RL} with \ac{TV} have more spread out errors and a bias on negative values. \ac{NIFTy} and \ac{RL} Tikhonov seem to have comparable statistics. RL with Tikhonov is faster than NIFTy, but less flexible for further applications.

\begin{figure*}
  \includegraphics[width=0.95\textwidth]{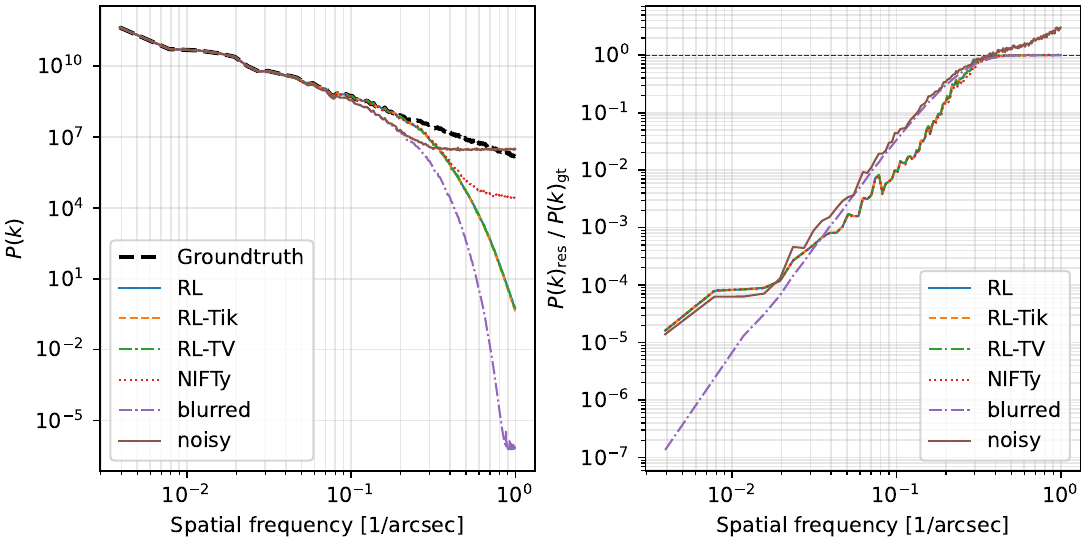}
    \caption{Power spectra (left panel) and the ratio of the power spectra of the residuals and the power spectrum of the ground truth (right panel) for image A. This shows the power spectra for the various methods in comparison with the power spectrum of the ground truth. 
    The black dashed line at $P_\mathrm{res}(k)/P_\mathrm{gt}(k) = 1$ marks the threshold above which algorithms assign more power to a given scale than is present in the ground truth. Crossing this line indicates overfitting.}
    \label{fig.:pspecA}
\end{figure*}
\begin{figure*}
  \includegraphics[width=0.95\textwidth]{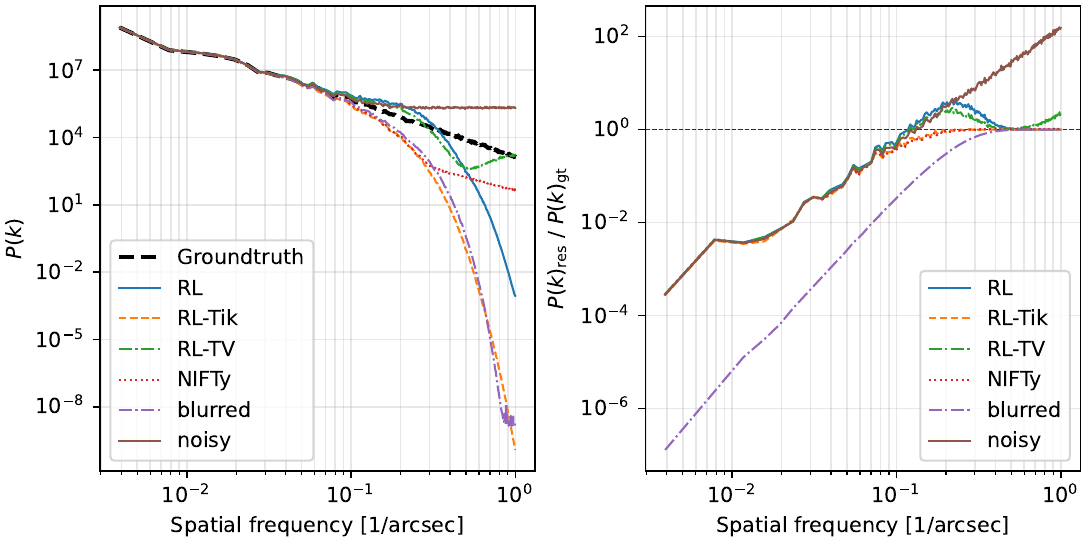}
  \caption{As Fig.~\ref{fig.:pspecA}, but for reconstruction image B.}
  \label{fig.:pspecB}
\end{figure*}
Figs.~\ref{fig.:pspecA} and \ref{fig.:pspecB} show the power spectra of the ground truth images A and B as well as the spectra of the blurred image, the blurred and noisy image and the reconstructions done with the various algorithms.
We calculate the power spectrum of a field $f$ with
\begin{equation}
P(k) = \frac{1}{N_k} \sum_{\lvert\mathbf{k}\rvert=k} \lvert\hat{f}(\mathbf{k})\rvert^2 \ ,
\end{equation}
where $\hat{f}$ is the two-dimensional Fourier transform of $f$,
\begin{equation}
  \hat{f}(\mathbf{k}) = \sum_{x,y} f(x,y) \cdot e^{-2\pi i (k_x x + k_y y)} \, .
\end{equation}
Here, $k$ is the length of the $\mathbf{k}$ vector, $k = \sqrt{k_x^2 + k_y^2}$ and $N_k$ is the number of pixels with that radius $k$. In this discussion, we evaluate the power spectrum for $f \in \{s, \epsilon \}$, i.e., the signal $s$ and the error $\epsilon$.

In Figs.~\ref{fig.:pspecA} and~\ref{fig.:pspecB}, we see that blurring by the \ac{PSF} removes power on short scales. 
We also observe that the noisy image, after drawing Poisson counts, gains a lot of power on these short scales. In the low-\ac{S/N} case of image B, the noise power even exceeds the original power.
In Fig.~\ref{fig.:pspecA}, we can see how well the different algorithms restore power on different scales. All methods recover some power on small scales. However, \ac{NIFTy} recovers more than the other algorithms.
In Fig.~\ref{fig.:pspecB}, the situation is slightly different. \ac{RL} with Tikhonov regularization restores some power and does not overfit. Plain \ac{RL} and \ac{RL} with \ac{TV}, however, overfit the data, meaning the "restored" power is due to fitted shot noise. We observe a clear bump at $k = 0.1-0.4 \, \mathrm{arcsec}^{-1}$ for these methods, at scales where the signal response falls below the noise level. 

The relative behavior between the power spectra can be observed even better in the right panels. 
These show the ratio $P_\mathrm{res}(k)/P_\mathrm{gt}(k)$, where $P_\mathrm{res}(k)$ is the power spectrum of the error $\epsilon$ and $P_\mathrm{gt}(k)$ is the power spectrum of the ground truth. A conservative algorithm should assign zero power to Fourier scales where no reliable information is available due to low \ac{S/N}. This yields $P_\mathrm{res}(k) = P_\mathrm{gt}(k)$, and therefore, $P_\mathrm{res}(k)/P_\mathrm{gt}(k) = 1$. Any ratio exceeding unity is therefore a clear indicator of overfitting on these scales, as the power of the error exceeds the power of the ground truth. 

\begin{figure}
  \includegraphics{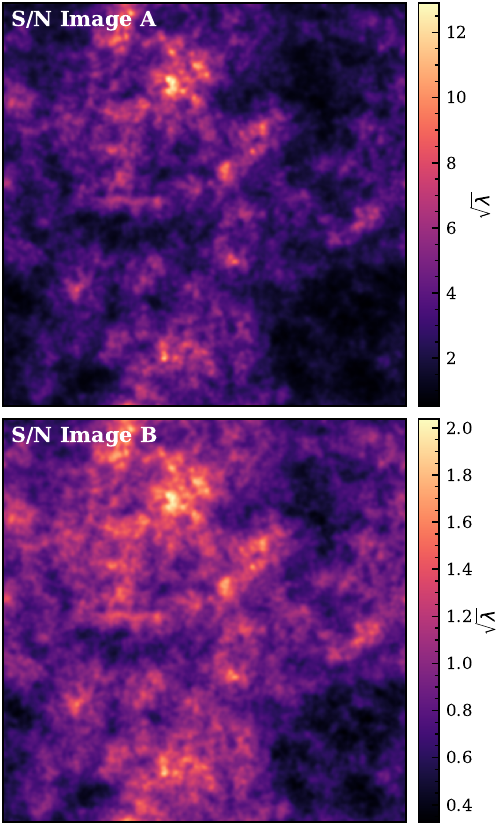}
  \caption{\ac{S/N} maps for image A and image B. The \ac{S/N}, defined as $\sqrt{\lambda_i}$ per pixel, ranges from 0.53 to 12.90 for image A and from 0.33 to 2.04 for image B, reflecting the significantly lower flux level of image B.}\label{fig.:snrmaps}
\end{figure}
\begin{figure}[h!]
  \includegraphics[width=0.49\textwidth]{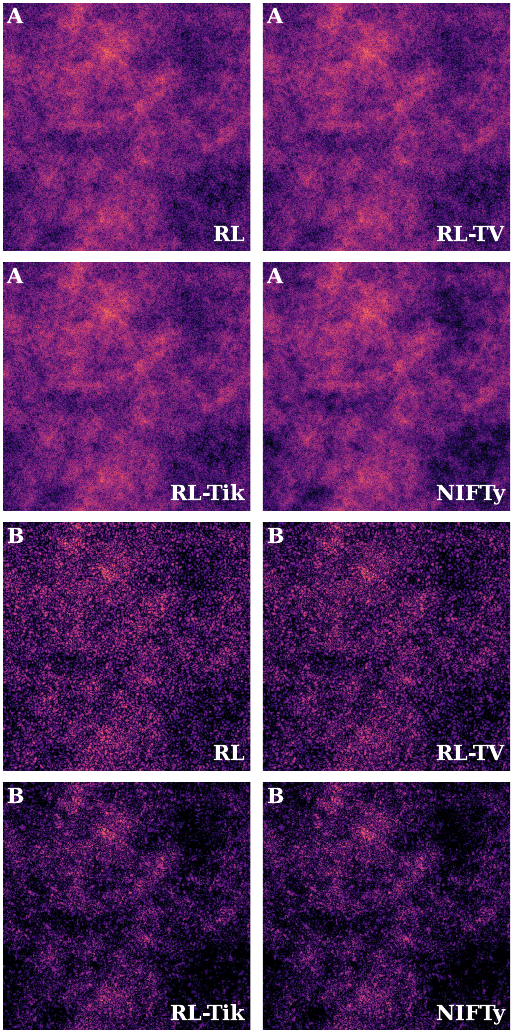}
  \caption{Absolute residuals for different methods, using  the same color scale as in Fig.~\ref{fig.:comparison_methods}.}\label{fig.:comparison_methods_residuals}
\end{figure}

In Fig.~\ref{fig.:comparison_methods_residuals}, we see the absolute residuals between ground truth and the individual methods used for deconvolution. 
For image A we can see that the residuals are comparable across all four methods. The largest differences occur in low-flux regions, e.g. lower-right or upper-right corners (see~\ref{fig.:snrmaps} for \ac{S/N}-maps). In these regions, all RL algorithms leave some small-scale residuals, while NIFTy appears to remove them.
For image B, we observe similar behavior, which is tied to the lower \ac{S/N}. The overfitting-to-noise behavior described in the power spectra discussion can be observed directly here as well.

\clearpage
\section{PSF approximation}\label{app.:psfapproximation}
\begin{figure}[h!]
  \includegraphics[width=0.49\textwidth]{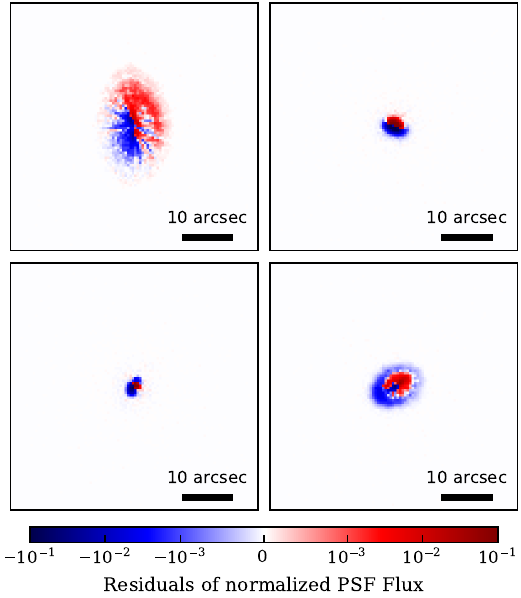}
  \caption{Difference plots between the simulated and the approximated \acp{PSF} from Fig.~\ref{fig.:psf_approx}. The colormap is linear between $\pm 10^{-3}$ and logarithmic elsewhere.}\label{fig.:psf_approx_diff}
\end{figure}

The \ac{PSF} approximation using the linearly interpolated patch-based approach yields the discussed results. 
To highlight these differences, we show them on a sym-log scale in Fig.~\ref{fig.:psf_approx_diff}. A sym-log scale is logarithmic except for a region around zero where it is linear, here from $-10^{-3}$ to $10^{-3}$.

\end{appendix}
\end{document}